\documentclass[12pt]{article}
\usepackage{amsmath,amssymb}
\usepackage{bm}
\usepackage{graphicx,color}
\usepackage{wrapfig}
\begin{document}
\title{
\begin{flushright}
\ \\*[-80pt] 
\begin{minipage}{0.2\linewidth}
\normalsize
\end{minipage}
\end{flushright}
{\Large \bf Sensitivity of the squark flavor mixing
 to the CP violation of  $K$, $B^0$ and $B_s$ mesons
\\*[20pt]}}

\author{
\centerline{Yusuke~Shimizu$^{1,}$\footnote{E-mail address:
 yusuke.shimizu@mpi-hd.mpg.de}, \ \ 
Morimitsu~Tanimoto$^{1,}$\footnote{E-mail address: tanimoto@muse.sc.niigata-u.ac.jp}, \ \  and \ \
Kei~Yamamoto$^{2,}$\footnote{E-mail address: yamamoto@muse.sc.niigata-u.ac.jp}}
\\*[20pt]
\centerline{
\begin{minipage}{\linewidth}
\begin{center}
$^1${\it \normalsize Max-Planck-Institute f\"ur Kernphysik,
Postfach 103980, D-69029 Heidelberg, Germany}
\\*[4pt]
$^2${\it \normalsize
Department of Physics, Niigata University,~Niigata 950-2181, Japan }
\\*[4pt]
$^3${\it \normalsize
Graduate~School~of~Science~and~Technology,~Niigata~University, \\ 
Niigata~950-2181,~Japan }
\end{center}
\end{minipage}}
\\*[70pt]}

\date{
\centerline{\small \bf Abstract}
\begin{minipage}{0.9\linewidth}
\vskip  1 cm
\small
We study  the sensitivity of  the squark flavor mixing
 to the CP violating phenomena of   $K$, $B^0$ and $B_s$ mesons
 in the framework of the split-family scenario, where the first and second family
 squarks are very heavy, ${\cal O}(10)$~TeV, on the other hand, the third family
  squark masses are at  ${\cal O}(1)$~TeV.
 In order to constrain the gluino-sbottom-quark mixing parameters,
 we input the experimental data of the CP violations of $K$, $B^0$, and $B_s$ mesons,
  that is $\epsilon_K$, $\phi_d$, and  $\phi_s$.
 The experimental upper bound of the chromo-EDM of the strange quark is also input.
 In addition,  we take account of the
 observed values $\Delta M_{B^0}$, $\Delta M_{B_s}$, the CKM mixing $|V_{ub}|$,
 and the branching ratio of $b\to s\gamma$.
 The allowed region of the mixing parameters are obtained as
$|\delta_{13}^{dL(dR)}|=0\sim 0.01$ and $|\delta_{23}^{dL(dR)}|=0\sim 0.04$.
By using these values, the deviations from the SM  are estimated 
in the CP violations of the $B^0$ and $B_s$ decays.
The deviation from the SM one is tiny in the CP asymmetries of $B^0\to \phi K_S$ and $B^0\to \eta 'K^0$ due to the  chromo-EDM of the strange quark.
On the other hand,  the  CP asymmetries $B_s \to \phi \phi$ and 
 $B_s \to \phi \eta '$ could be largely deviated from the SM predictions. 
We also  predict the time dependent CP asymmetry of
$B^0\to K^0\bar K^0$  and 
the semi-leptonic CP asymmetries 
of  $B^0 \to \mu ^-X$ and   $B_s \to \mu ^-X$.
 We expect those  precise measurements  at Belle II,
 which  will provide us
interesting tests for the squark flavor mixing.
\end{minipage}
}

\begin{titlepage}
\maketitle
\thispagestyle{empty}
\end{titlepage}

\section{Introduction}
\label{sec:Intro}

The flavor physics is on the new stage in the light of LHCb data.
The LHCb collaboration has reported 
new data of the CP violation of the $B_s$ meson and the branching ratios 
of rare $B_s$ decays~\cite{Bediaga:2012py}-\cite{LHCb:2011ab}.
For many years the CP violation in the $K$ and $B^0$ mesons 
has been successfully understood within the framework of the standard model (SM), 
so called Kobayashi-Maskawa (KM) model, 
where the source of the CP violation is the KM phase 
in the quark sector with three families. 
However, the new physics has been expected to be indirectly discovered
in the precise data of $B^0$ and $B_s$ meson decays at the LHCb experiment and the further coming  experiment, Belle II. 

The supersymmetry (SUSY) is one of the most attractive candidates for the new physics. 
The SUSY signals have not been observed yet  
although the Higgs-like events have been confirmed~\cite{Higgs}. 
Since the lower bounds of the superparticle masses increase gradually, 
the squark and the gluino masses are supposed  
to be at the TeV scale~\cite{squarkmass}. 
While, there are new sources of the CP violation if the SM is 
extended to the SUSY models. The soft squark mass matrices contain 
the CP-violating phases, which contribute to the flavor changing 
neutral current (FCNC) with the CP violation. 
Therefore, we expect the effect of the SUSY contribution in the CP-violating phenomena. 
However, the clear deviation from the  SM prediction
has not been observed yet in the LHCb experiment~\cite{Bediaga:2012py}-\cite{LHCb:2011ab}.

The LHCb collaboration presented the time dependent CP asymmetry 
in the non-leptonic $B_s\to {J/\psi \phi}$ decay \cite{LHCb:2011aa,LHCb:2011ab,Aaij:2013oba}, which gives a constraint of 
the SUSY contribution on the $b\to s$ transition. 
They have  also reported the first measurement of the CP violating phase 
in the $B_s \to \phi \phi$ decay \cite{Aaij:2013qha}. 
This decay process is occurred at the one-loop level in the SM, where the CP violating phase
is very small. On the other hand, the gluino-squark mediated flavor changing process 
provides new CP violating phases. Thus, the CP asymmetry of $B_s \to \phi \phi$
is expected to be deviated considerably from the SM one.
In this work, we discuss the sensitivity of the SUSY contribution to the CP asymmetry of $B_s\rightarrow \phi\phi$ and  $B_s\to \phi \eta '$ 
by taking account of constraints from other experimental data of the CP violation.
For these decay modes,  the most important process of the SUSY contribution  is the gluino-squark mediated 
flavor changing process \cite{King:2010np}- \cite{Shimizu:2012zw}.
This FCNC effect is constrained by the CP violations in  $B^0\to {J/\psi K_S}$ and
 $B_s\to {J/\psi \phi}$ decays. 
 The CP violation of $K$ meson, $\epsilon_K$,  also provides a severe constraint to the gluino-squark mediated FCNC. In the SM, $\epsilon_K$ is proportional to 
 $\sin (2\beta)$ which is derived from the time dependent CP asymmetry 
 in $B^0\to J/\psi K_s$  decay
\cite{Buras:2008nn}. 
 The relation between $\epsilon_K$ and  $\sin (2\beta)$ is examined 
 by taking account of  the gluino-squark mediated FCNC \cite{Mescia:2012fg}.
 

The time dependent CP asymmetry of 
$B^0\to \phi K_S$, $B^0\to \eta ' K^0$,  and  $B^0 \to K^0 \bar K^0$ decays
are also  attractive ones to search for the gluino-squark mediated FCNC 
because the penguin amplitude dominates this process as well as $B_s \to \phi \phi$.
Furthermore, we  discuss the FCNC with the CP violation
in the semileptonic CP asymmetries of $B^0$ and $B_s$ mesons.

In addition, it is remarked  that the upper-bound of the chromo-EDM(cEDM) of the strange quark gives a severe constraint for the gluino-squark mediated $b\to s$ transition
\cite{Hisano:2003iw}-\cite{Fuyuto:2012yf}.
   

 The lower bounds of the squark masses increase gradually. 
The gluino mass is  expected 
to be larger than $1.3$~TeV, and the squarks of the first and second
 families are also heavier than $1.4$~TeV~\cite{squarkmass}. 
 Therefore, we take the split-family scenario, in which the first and second family
 squarks are very heavy, ${\cal O}(10)$~TeV, while the third family
  squark masses are at  ${\cal O}(1)$~TeV.
  Then,  the  $s\to d$ transition mediated by the first and second family squarks
  is naturally suppressed by their heavy masses, and competing process is mediated by the second order contribution of the third family squark.
  In order to estimate the gluino-squark mediated FCNC 
  for the  $K$, $B^0$ and $B_s$ meson decays comprehensively,
  we work in the basis of  the squark mass eigenstate. 
Then, the $6\times 6$ mixing matrix among down-squarks and down-quarks is studied
 by input of the experimental constraints.

In section 2, we present the formulation of the  gluino-squark mediated
transition in our  split-family scenario.
In section 3, we discuss the gluino-squark mediated FCNC contribution to  $\epsilon_K$.
In section 4, we discuss the sensitivity of the gluino-squark mediated FCNC to the CP violation of  the non-leptonic  and the semi-leptonic decays of $B^0$ and $B_s$ 
mesons.
Section 5 is devoted to the summary. 


\section{ CP violation through squark flavor mixing}
\label{sec:Deviation}
\subsection{Squark flavor mixing}
Let us discuss the gluino-squark mediated flavor changing process 
as the dominate SUSY contribution. We give the $6\times 6$ squark mass matrix to be $M_{\tilde q}$
$(\tilde q=\tilde u, \tilde d)$  in the super-CKM basis. In order to go to the  diagonal basis of the squark mass matrix, we rotate $M_{\tilde q}$ as
\begin{equation}
 \tilde m_{\tilde q\rm dia}^2=\Gamma _{G}^{(q)}M_{\tilde q}^2 \Gamma _{G}^{(q)  \dagger} \ ,
\end{equation}
where  $\Gamma _{G}^{(q)}$ is the $6\times 6$ unitary matrix, and
we decompose it into the  $3\times 6$ matrices 
 as $\Gamma _{G}^{(q)}=(\Gamma _{GL}^{(q)}, \ \Gamma _{GR}^{(q))})^T$ in the following expressions. 
Then, the gluino-squark-quark interaction is given as
 \begin{equation}
\mathcal{L}_\text{int}(\tilde gq\tilde q)=-i\sqrt{2}g_s\sum _{\{ q\} }\widetilde q_i^*(T^a)
\overline{\widetilde{G}^a}\left [(\Gamma _{GL}^{(q)})_{ij}{\bm L}
+(\Gamma _{GR}^{(q)})_{ij}{\bm R}\right ]q_j+\text{h.c.}~,
\end{equation}
where $\widetilde G^a$ denotes the gluino field, and  ${\bm L}$ and ${\bm R}$ are projection operators. 
%
This interaction leads to the gluino-squark mediated flavor changing process 
with $\Delta F=2$ and  $\Delta F=1$ 
through the  box and  penguin diagrams.

 In our framework, the squarks of the first and second families are heavier than
  multi-TeV, on the other hand, the masses of the third family squarks, stop and sbottom, are around  $1$~TeV. Therefore, the first and second squark contribution is  suppressed
in the gluino-squark mediated flavor changing process by their heavy masses. 
 The stop and sbottom interactions dominate the gluino-squark mediated flavor changing process. 
Then, the sbottom interaction dominates  $\Delta B=2$ and  $\Delta B=1$ processes.
 We take a suitable parametrizations of
 $\Gamma _{GL}^{(d)}$ and $\Gamma _{GR}^{(d)}$ as follows \cite{Mescia:2012fg}:
 \begin{align}
\Gamma _{GL}^{(d)}&=
\begin{pmatrix}
1 & 0 & \delta _{13}^{dL}c_\theta & 0 & 0 & -\delta _{13}^{dL}s_\theta e^{i\phi } \\
0 & 1 & \delta _{23}^{dL}c_\theta & 0 & 0 & -\delta _{23}^{dL}s_\theta e^{i\phi } \\
-{\delta _{13}^{dL}}^* & -{\delta _{23}^{dL}}^* & c_\theta & 0 & 0 & -s_\theta e^{i\phi }
\end{pmatrix}, \nonumber \\
\nonumber\\
\Gamma _{GR}^{(d)}&=
\begin{pmatrix}
0 & 0 & \delta _{13}^{dR}s_\theta e^{-i\phi } & 1 & 0 & \delta _{13}^{dR}c_\theta \\
0 & 0 & \delta _{23}^{dR}s_\theta e^{-i\phi } & 0 & 1 & \delta _{23}^{dR}c_\theta \\
0 & 0 & s_\theta e^{-i\phi } & -{\delta _{13}^{dR}}^* & -{\delta _{23}^{dR}}^* & c_\theta 
\end{pmatrix},
\label{mixing}
\end{align}
where $c_\theta =\cos \theta$ and $s_\theta =\sin \theta$, 
with the mixing angle  $\theta$ in the $\tilde b_{L,R}$ sector 
and $\delta _{j3}^{dL}$, $\delta _{j3}^{dR}$ are the couplings responsible for the flavor transitions. 
By using these rotation matrices, we estimate the gluino-sbottom mediated flavor changing amplitudes in the  $K$, $B^0$, and $B_s$ meson decays.

For the numerical analysis, we fix  sbottom masses.
The third family squarks can have substantial mixing between the left-handed squark and the right-handed one due to large Yukawa couplings.
 In our numerical calculation, we take the typical mass eigenvalues $m_{\tilde b_1}$  and  $m_{\tilde b_2}$, and the gluino mass $m_{\tilde g}$ as follows:
  \begin{equation}
  m_{\tilde b_1}=1 \ {\rm TeV}, \qquad m_{\tilde b_2}=1.1 \ {\rm TeV},
  \qquad m_{\tilde g}=2 \ {\rm TeV},
 \end{equation}
  where we take account of the present experimental bounds \cite{squarkmass}.
  Then, we can roughly estimate the mixing angle $\theta$  between the left-handed sbottom and the right-handed one by using RGE's under the assumption of the universal mass
 of the GUT scale although it depends on the SUSY parameters in details 
\cite{Martin:1997ns}. 
Therefore, we scatter  the left and right mixing  $\theta$ in the range of $10^\circ-35^\circ$
 in our numerical calculations.
 The mixing parameters $\delta _{13}^{dL}$ and 
 $\delta _{23}^{dL}$ are complex, and  will be constrained  by the experimental data.
 For simplicity, we take
 \begin{equation}
  |\delta _{13}^{dR}|=|\delta _{13}^{dL}|, \qquad
|\delta_{23}^{dR}|=|\delta_{23}^{dL}|,
 \end{equation}
 on the other hand, the phases of $\delta_{23}^{dR}$and $\delta_{1 3}^{dR}$, and  the phase $\phi$ are   free parameters.
Therefore, we have three mixing angles and five  phases
 in the mixing matrices of Eq.(\ref{mixing}), which are free parameters in our calculations.
 
\subsection{CP violation in $\Delta B=2$ and $\Delta B=1$ processes}
 Let us discuss the SUSY contribution in the $\Delta B=2$ process.
The contribution of new physics to the dispersive part $M_{12}^q$ 
is parameterized as 
\begin{equation}
M_{12}^q=M_{12}^{q,\text{SM}}+M_{12}^{q,\text{SUSY}}=
M_{12}^{q,\text{SM}}(1+h_qe^{2i\sigma _q})~, \quad (q=d,s)
\end{equation}
where $M_{12}^{q,\text{SM}}$ and $M_{12}^{q,\text{SUSY}}$ are the SM and the SUSY
contributions.
 The  parameters  $h_q$ and $\sigma_q$ are given in terms of mixing parameters
 of Eq.(\ref{mixing}).
 The $M_{12}^{q,\text{SUSY}}$ are given explicitly in Appendix A. 
By inputting experimental data of $\epsilon_K$, $\Delta M_{B^0}$, $\Delta M_{Bs}$
, $\sin (2\beta)$, and $\sin (2\beta_s)$, we constrain  the magnitude $h_q$ and the phase $\sigma_q$. 

The indirect CP violation leads to the non-zero asymmetry $a_{sl}^q$
 in the semileptonic decays  $B_q\to \mu ^-X(q=d,s)$ with "wrong-sign" such as: 
\begin{equation}
a^q_{sl}\equiv \frac{\Gamma (\bar B_q\rightarrow \mu ^+X)-\Gamma (B_q\rightarrow \mu ^-X)}
{\Gamma (\bar B_q\rightarrow \mu ^+X)+\Gamma (B_q\rightarrow \mu ^-X)}
\simeq \text{Im} \left (\frac{\Gamma _{12}^q}{M_{12}^q}\right )=
\frac{|\Gamma _{12}^q|}{|M_{12}^q|}\sin \phi^q_{sl},
\end{equation}
where $\Gamma _{12}^q$ is  the absorptive part in the effective Hamiltonian of 
the $B_q$-$\bar B_q$ system, where $B_d$ is denoted as the  $B^0$ meson
in this paper.
The SM contribution to the absorptive part $\Gamma _{12}^q$ is dominated 
by tree-level decay $b\to c\bar c s$ etc..
Therefore, we assume $\Gamma _{12}^q=\Gamma _{12}^{q,\text{SM}}$ in our calculation.
 In the SM, the CP phases are read  \cite{Lenz:2011ti},
 \begin{equation}
  \phi^{s\rm SM}_{sl}=(3.84\pm 1.05)\times 10^{-3},  \qquad
  \phi^{d\rm SM}_{sl}=-(7.50\pm 2.44)\times 10^{-2},
 \end{equation}
which correspond to 
\begin{equation}
  a^{s\rm SM}_{sl}=(1.9\pm 0.3)\times 10^{-5},  \qquad
  a^{d\rm SM}_{sl}=-(4.1\pm 0.6)\times 10^{-4}.
  \label{semiCP}
 \end{equation}
 The recent experimental data of these asymmetries are given as 
\cite{Vesterinen:2013jia,PDG}
 \begin{equation}
  a^{s}_{sl}=(-0.24\pm 0.54\pm 0.33)\times 10^{-2},  \qquad
  a^{d}_{sl}=(-0.3\pm 2.1)\times 10^{-3}.
  \label{semiexp}
 \end{equation}

 
  There are many interesting  non-leptonic CP violating decays
to search for new physics.
 The effective Hamiltonian for the $\Delta B=1$ process  is given as follows:
\begin{equation}
H_{eff}=\frac{4G_F}{\sqrt{2}}\left [\sum _{q'=u,c}V_{q'b}V_{q'q}^*
\sum _{i=1,2}C_iO_i^{(q')}-V_{tb}V_{tq}^*
\sum _{i=3-6,7\gamma ,8G}\left (C_iO_i+\widetilde C_i\widetilde O_i\right )\right ],
\label{hamiltonian}
\end{equation}
where $q=s,d$. The local operators are given as 
\begin{align}
&O_1^{(q')}=(\bar q_\alpha\gamma _\mu P_Lq_\beta')
(\bar q_\beta'\gamma ^\mu P_Lb_\alpha),
\qquad O_2^{(q')}=(\bar q_\alpha\gamma _\mu P_Lq_\alpha')
(\bar q_\beta'\gamma ^\mu P_Lb_\beta), \nonumber \\
&O_3=(\bar q_\alpha\gamma _\mu P_Lb_\alpha)\sum _Q(\bar Q_\beta\gamma ^\mu P_LQ_\beta),
\quad O_4=(\bar q_\alpha\gamma _\mu P_Lb_\beta)\sum _Q(\bar Q_\beta\gamma ^\mu P_LQ_\alpha), \nonumber \\
&O_5=(\bar q_\alpha\gamma _\mu P_Lb_\alpha)\sum _Q(\bar Q_\beta\gamma ^\mu P_RQ_\beta),
\quad O_6=(\bar q_\alpha\gamma _\mu P_Lb_\beta)\sum _Q(\bar Q_\beta\gamma ^\mu P_RQ_\alpha), \nonumber \\
&O_{7\gamma }=\frac{e}{16\pi ^2}m_b\bar q_\alpha\sigma ^{\mu \nu }P_Rb_\alpha
F_{\mu \nu }, 
\qquad O_{8G}=\frac{g_s}{16\pi ^2}m_b\bar q_\alpha\sigma ^{\mu \nu }
P_RT_{\alpha\beta}^ab_\beta G_{\mu \nu }^a,
\end{align}
where $P_R=(1+\gamma _5)/2$, $P_L=(1-\gamma _5)/2$, and $\alpha $, $\beta $ are color 
indices, and $Q$ is taken to be $u,d,s,c$ quarks. 
Here, $C_i$'s and $\widetilde C_i$'s are the Wilson coefficients at the relevant mass scale, 
and $\widetilde O_i$'s are the operators by replacing $L(R)$ with $R(L)$ 
in $O_i$. In this paper, $C_i$ includes both SM contribution and squark-gluino one, 
such as $C_i=C_i^{\rm SM}+C_i^{\tilde g}$, where 
$C_i^{\text{SM}}$'s are given in Ref.~\cite{Buchalla:1995vs}. 
The Wilson coefficients of the gluino-squark contribution 
$C_{7\gamma}^{\tilde g}$ and $C_{8G}^{\tilde g}$ are presented in Appendix B,
 where it is remarked that the magnitudes of 
$C_{7\gamma}^{\tilde g}(m_b)$ and $C_{8G}^{\tilde g}(m_b)$ are reduced
by the cancellation between the contributions of two sbottom
 $\tilde b_1$ and $\tilde b_2$.
 
The Wilson coefficients of $C_{7\gamma}^{\tilde g}(m_b)$ and 
$C_{8G}^{\tilde g}(m_b)$ at the $m_b$ scale are given at the leading order of QCD as follows~\cite{Buchalla:1995vs}: 
\begin{equation}
\begin{split}
C_{7\gamma}^{\tilde g}(m_b)
&= \zeta C_{7\gamma}^{\tilde g}(m_{\tilde g})
+\frac{8}{3}(\eta-\zeta) C_{8G}^{\tilde g}(m_{\tilde g}), \cr
C_{8G}^{\tilde g}(m_b)
&=\eta C_{8G}^{\tilde g}(m_{\tilde g}),
\end{split}
\end{equation}
where 
\begin{equation}
\zeta=\left ( 
 \frac{\alpha_s(m_{\tilde g})}{\alpha_s(m_t)} \right )^{\frac{16}{21}}
 \left ( 
 \frac{\alpha_s(m_t)}{\alpha_s(m_b)} \right )^{\frac{16}{23}} \ , \qquad
 \eta=\left ( 
 \frac{\alpha_s(m_{\tilde g})}{\alpha_s(m_t)} \right )^{\frac{14}{21}}
 \left ( 
 \frac{\alpha_s(m_t)}{\alpha_s(m_b)} \right )^{\frac{14}{23}} \ .
 \end{equation}

Let us discuss the time dependent CP asymmetries of $B^0$ and $B_s$ decaying 
into the final state $f$, which are defined as~\cite{Aushev:2010bq} 
\begin{equation}
S_f=\frac{2\text{Im}\lambda _{f}}{1+|\lambda_{f}|^2}\ ,
\label{sf}
\end{equation}
where 
\begin{equation}
\lambda_{f}=\frac{q}{p} \bar \rho\ , \qquad 
\frac{q}{p}\simeq \sqrt{\frac{M_{12}^{q*}}{M_{12}^{q}}}, \qquad 
\bar \rho \equiv 
\frac{\bar A(\bar B_q^0\to f)}{A(B_q^0\to f)}.
\label{lambdaf}
\end{equation}
Here $M_{12}^q(q=s,d)$ include
the SUSY contribution in addition to the SM one.

In the $B^0\to J/\psi  K_S$ and $B_s\to J/\psi  \phi$ decays,
 we write $\lambda_{J/\psi  K_S}$ and $\lambda_{J/\psi  \phi}$
in terms of phase factors, respectively:
\begin{equation}
\lambda_{J/\psi  K_S}\equiv 
-e^{-i\phi _d}, \qquad \lambda _{J/\psi \phi } \equiv e^{-i\phi _s}.
\label{new}
\end{equation}
In the SM, the angle $\phi_d$ is given as $\phi_d=2\beta$, in which $\beta$ is 
one angle of  the unitarity triangle with respect to $B^0$.
On the other hand,   $\phi_s$ is given as $\phi_s=-2\beta_s$,
in which $\beta_s$ is one angle  of  the unitarity triangle   for $B_s$.
Once $\phi_d$ is input, the SM predicts $\phi_s $ as \cite{Charles:2004jd}
\begin{equation}
\phi_s=-0.0363\pm 0.0017\ .
\end{equation}
The recent experimental data of these phases are \cite{Aaij:2013oba,Amhis:2012bh}
\begin{equation}
\sin \phi _d=0.679\pm 0.020\ , \qquad \phi_s=0.07\pm 0.09\pm 0.01 \ ,
\label{phasedata}
\end{equation}
 in which the contribution of the gluino-squark-quark interaction is expected to be found
 because of 
 \begin{equation}
\phi _d=2\beta+\arg(1+h_d e^{2 i\sigma_d}) \ , 
\qquad \phi_s=-2\beta_s+\arg(1+h_s e^{2 i\sigma_s}) \ ,
\label{phiSUSY}
\end{equation}
where $\beta(\beta_s)$ is given in terms of the CKM matrix elements.
 These experimental values also constrain the mixing parameters in Eq.(\ref{mixing}).

Let us consider the contribution from the gluino-sbottom-quark interaction
 in the nonleptonic decays of the $B^0$ meson.
Since the $B^0\to J/\psi K_S$ process occurs at the tree level in  the SM, 
the CP asymmetry in this process mainly originates from  $M_{12}^d$.
The CP asymmetries of the  penguin dominated decays $B^0\to \phi K_S$ and $B^0\to\eta 'K^0$ 
also come from  $M_{12}^d$ in the SM. 
Then, the CP asymmetries of
 $B^0\to J/\psi K_S$,  $B^0\to \phi K_S$, and 
$B^0\to \eta 'K^0$ decays are expected to be the same magnitude within $10 \%$.
On the other hand, if the gluino-sbottom-quark interaction contributes to the decay 
at the one-loop level, its magnitude could be  comparable 
to the SM penguin one 
in  $B^0\to \phi K_S$ and $B^0\to \eta 'K^0$ decays, 
but the effect of the gluino-sbottom-quark interaction  is tiny in the $B^0\to J/\psi K_S$ 
decay because this process is at the tree level in the SM. 
Therefore, there is a possibility to 
find the SUSY contribution  by  observing 
the different CP asymmetries among those processes~\cite{Khalil:2003bi,Endo:2004dc}.
 
The time dependent CP asymmetry $S_{J/\psi K_S}$ has been precisely measured.
We take the data of 
these time dependent CP asymmetries in HFAG~\cite{Amhis:2012bh}, which are 
\begin{equation}
S_{ J/\psi K_S}=0.679\pm 0.020 \ , \qquad 
S_{\phi K_S}= 0.74^{+0.11}_{-0.13}\ , \qquad 
S_{\eta 'K^0}= 0.59\pm 0.07\ .
\label{Sfdata}
\end{equation}
These values may be regarded to be  same within the experimental error-bar. 
Thus, the experimental values are consistent with the prediction of the SM. 
In other words, 
these data severely may constrain the flavor mixing parameter
 $\delta^{dL(dR)}_{23}$. 
 
 Recently, LHCb reported the first
flavor-tagged measurement of the time-dependent CP-violating asymmetry
in the $B_s$  decay \cite{Aaij:2013qha}. In this decay process, the CP-violating weak phase
arises due to the CP violation in the interference between $B_s-\bar B_s$ mixing and 
the $b\to s\bar ss$ gluonic penguin decay amplitude. 
The CP-violating phase $\phi_s$ is measured to be in the interval 
\begin{equation}
\phi_s=[-2.46,-0.76]\ \  {\rm rad}  \ ,
\label{phiphidata} 
\end{equation}
at 68\%C.L. \cite{Aaij:2013qha}.
We expect that the precise data  will be presented in the near future.

 \subsection{The $b\to s$ transition}

The CP asymmetries $S_f$ for $B^0\to \phi K_S$ and $B^0\to \eta 'K^0$ 
are given in terms of $\lambda_f$ in Eq.~(\ref{lambdaf}):
\begin{align}
\lambda _{\phi K_S,~\eta 'K^0}&=-e^{-i\phi _d}\frac{\displaystyle \sum _{i=3-6,7\gamma ,8G}
\left (C_i^\text{SM}\langle O_i \rangle +C_i^{\tilde g}\langle O_i \rangle +
\widetilde C_i^{\tilde g}\langle \widetilde O_i \rangle \right )}
{\displaystyle \sum _{i=3-6,7\gamma ,8G}\left (C_i^{\text{SM}*}\langle O_i \rangle 
+C_i^{{\tilde g}*}\langle O_i \rangle +\widetilde C_i^{{\tilde g}*}\langle \widetilde O_i 
\rangle \right )}~,
\label{asymBd}
\end{align}
where $\langle O_i \rangle $ is the abbreviation of $\langle f|O_i|B^0\rangle $. 
It is noticed $\langle \phi K_S|O_i|B^0\rangle =\langle \phi K_S|\widetilde O_i|B^0\rangle $ 
and $\langle \eta 'K^0|O_i|B^0\rangle =-\langle \eta 'K^0|\widetilde O_i|B^0\rangle $, 
because these final states have different parities~\cite{Endo:2004dc,Khalil:2003bi}. 
Since the dominant term comes from the gluon penguin $C_{8G}^{\tilde g}$, 
the decay amplitudes of $f=\phi K_S$ and $f=\eta 'K^0$ are given as follows: 
\begin{align}
\bar A(\bar B^0 \to \phi K_S)
& \propto C_{8G}(m_b) + {\tilde C}_{8G}(m_b), \nonumber \\
\bar A(\bar B^0 \to \eta '\bar K^0)
& \propto C_{8G}(m_b) - {\tilde C}_{8G}(m_b).
\end{align}
Since ${\tilde C}_{8G}(m_b)$ is suppressed compared to $C_{8G}(m_b)$ in the SM, 
the magnitudes of the time dependent CP asymmetries 
$S_f \ (f=J/\psi \phi, \ \phi K_S,\  \eta 'K^0)$ are almost same in the SM prediction. 
However, the squark flavor mixing gives the unsuppressed ${\tilde C}_{8G}(m_b)$, 
then, the CP asymmetries in those decays are expected to be deviated among them. 
Therefore, those experimental data  give us the tight constraint for $C_{8G}(m_b)$ and ${\tilde C_{8G}}(m_b)$. 

We have also $\lambda _{f}$ for $B_s\to \phi \phi $ and $B_s\to \phi \eta '$ as follow: 
\begin{align}
\lambda _{\phi \phi ,\phi \eta '}&=e^{-i\phi _s}\frac{\displaystyle \sum _{i=3-6,7\gamma ,8G}
C_i^\text{SM}\langle O_i \rangle +C_i^{\tilde g}\langle O_i \rangle +
\widetilde C_i^{\tilde g}\langle \widetilde O_i \rangle }
{\displaystyle \sum _{i=3-6,7\gamma ,8G}
C_i^{\text{SM}*}\langle O_i \rangle +C_i^{{\tilde g}*}
\langle O_i \rangle +\widetilde C_i^{{\tilde g}*}\langle \widetilde O_i \rangle }~,
\label{asymBs}
\end{align}
with $\langle \phi \phi |O_i|B_s\rangle =-\langle \phi \phi |\widetilde O_i|B_s\rangle $ 
and $\langle \phi \eta '|O_i|B_s\rangle =\langle \phi \eta '|\widetilde O_i|B_s\rangle $. 
The decay amplitudes of $f=\phi \phi $ and $f=\phi \eta '$ are given as follows: 
\begin{align}
\bar A(\bar B_s \to \phi \phi )
& \propto C_{8G}(m_b) - {\tilde C}_{8G}(m_b), \nonumber \\
\bar A(\bar B_s \to \phi \eta ')
& \propto C_{8G}(m_b) + {\tilde C}_{8G}(m_b).
\label{phi-eta}
\end{align}
Since  $C_{8G}\langle O_{8G}\rangle $ 
and $\tilde C_{8G}\langle \tilde O_{8G}\rangle $ 
dominate these amplitudes,  our numerical results are insensitive
 to the hadronic matrix elements.
In order to obtain precise results,
we also take account of the small contributions 
from other Wilson coefficients $C_i~(i=3,4,5,6)$ and $\tilde C_i~(i=3,4,5,6)$ 
in our calculations. 
We estimate each hadronic matrix element 
by using the factorization relations in Ref.~\cite{Harnik:2002vs}: 
\begin{equation*}
\langle O_{3} \rangle =\langle O_{4} \rangle =\left( 1+\frac{1}{N_c} \right) \langle O_{5} \rangle,
\quad \langle O_{6} \rangle =\frac{1}{N_c}\langle O_{5} \rangle,
\end{equation*}
\begin{equation}
\langle O_{8G} \rangle =\frac{\alpha _s(m_b)}{8 \pi }
\left( -\frac{2 m_b}{ \sqrt{\langle q^2 \rangle }}\right )
\left( \langle O_4 \rangle +\langle O_6 \rangle -\frac{1}{N_c}(\langle O_3 \rangle 
+\langle O_5 \rangle )\right ), 
\end{equation}
where $\langle q^2 \rangle ={\rm 6.3~GeV^2}$ 
and $N_c=3$ is the number of colors. 
One may worry about the reliability of these naive factorization relations. 
However, this approximation has been justified numerically 
in the relevant $b\to s$ transition as seen in the calculation of PQCD~\cite{Mishima:2003wm}. 

\subsection{The $b\to d$ transition}
\label{sec:bdtransitions}

The time dependent CP asymmetry $S_{K^0\bar K^0}$ in 
the $B^0\to K^0\bar K^0$ decay 
is also the interesting one to search for the new physics 
since there is no tree process of the SM in the $B^0\to K^0\bar K^0$ decay 
\cite{Giri:2004af,Fleischer:2004vu}. 
The amplitude $\bar A(\bar B^0\to K^0\bar K^0)$ 
is given in Ref.~\cite{Giri:2004af}, 
in which the QCD factorization is taken for the hadronic matrix elements~\cite{Muta:2000ti}
 ~\footnote{Improved analyses with $SU(3)$ flavor symmetry were presented in
 Refs.~\cite{DescotesGenon:2006wc, Baek:2005wx, Baek:2006pb}.}, as 
\begin{equation}
\bar A(\bar B^0\to K^0\bar K^0)
\simeq 
\frac{4G_F}{\sqrt{2}}\sum _{q=u,c}V_{qb}V_{qd}^*
\left [a_4^q(m_b)+r_\chi a_6^q(m_b)\right ]X.
\end{equation}
Here $X$ is the factorized matrix element (See Ref.~\cite{Giri:2004af}.) as 
\begin{equation}
X=-if_KF_0(m_K^2)(m_B^2-m_K^2),
\end{equation}
where $f_K$ and $F_0(m_K^2)$ denote the decay coupling constant of the 
$K$ meson and the form factor, respectively, 
and $r_\chi=2m_K^2/((m_b-m_s)(m_s+m_d))$ denotes the chiral enhancement factor. 
The coefficients $a_i^q$'s are given as~\cite{Giri:2004af,Muta:2000ti} 
\begin{align}
a_4^q(m_b)&=(C_4-\tilde C_4)+\frac{(C_3-\tilde C_3)}{N_c}+\frac{\alpha _s(m_b)}{4\pi }\frac{C_F}{N_c}
\Bigg [(C_3-\tilde C_3)\left [F_K+G_K(s_d)+G_K(s_b)\right ] \nonumber \\
&\hspace{1cm}+C_2G_K(s_q)+\left [ (C_4-\tilde C_4)+(C_6-\tilde C_6)\right ] 
\sum _{f=u}^bG_K(s_f)+(C_{8G}-\tilde C_{8G})G_{K,g}\Bigg ], \nonumber \\
a_6^q(m_b)&=(C_6-\tilde C_6)+\frac{(C_5-\tilde C_5)}{N_c}+\frac{\alpha _s(m_b)}{4\pi }\frac{C_F}{N_c}
\Bigg [(C_3-\tilde C_3)\left [G_K'(s_d)+G_K'(s_b)\right ] \nonumber \\
&\hspace{1cm}+C_2G_K'(s_q)+\left [(C_4-\tilde C_4)+(C_6-\tilde C_6)\right ]
\sum _{f=u}^bG_K'(s_f)+(C_{8G}-\tilde C_{8G})G_{K,g}'\Bigg ],
\label{coefficients-BKK}
\end{align}
where $q$ takes $u$ and $c$ quarks, $C_F=(N_c^2-1)/(2N_c)$, 
and the loop functions $F_K$, $G_K$, $G_{K,g}$, $G_K'$, 
and $G_{K,g}'$ are given in Refs.~\cite{Giri:2004af,Muta:2000ti}. 
The internal quark mass in the penguin diagrams 
enters as $s_f=m_f^2/m_b^2$.
\footnote{The $C_i^{\tilde g}$ 
in Eq.~(\ref{coefficients-BKK}) should be replaced with
 $ [(V_{tb}V_{tq}^*)/(V_{qb}V_{qd}^*) ]C_i^{\tilde q}$ 
in  Appendix B.} 
The minus sign in front of $\tilde C_i~(i=3-6,8G)$ comes from the 
parity of the final state. 
The CP asymmetry $S_{K^0\bar K^0}$ is  given  in terms of $\lambda _{K^0\bar K^0}$:
\begin{equation}
\lambda _{K^0\bar K^0}=-e^{-i\phi_d}\ 
\frac{\bar A(\bar B^0 \to K^0 \bar K^0)}{A(B^0\to K^0 \bar K^0)}.
\end{equation}
\subsection{Chromo EDM of strange quark}
In addition to the CP violating processes with  $\Delta B=2,\ 1$, we should discuss the T violation of flavor conserving process,
 that is the electric dipole moment.
The T violation is expected to be observed 
 in the electric dipole moment of the neutron and the electron.
The experimental upper bound of the  electric dipole moment of the neutron
provides us the upper-bound of  the chromo-EDM(cEDM) of the strange quark 
\cite{Hisano:2003iw}-\cite{Fuyuto:2012yf}. 

The cEDM of the strange quark $d_{s}^C$ is given 
in terms of the gluino-sbottom-quark interactions as seen in Appendix C.
The upper bound of the cEDM of the strange quark is given
by the experimental upper bound of the neutron EDM as \cite{Fuyuto:2012yf},
\begin{equation}
 e|d_s^C|<0.5\times 10^{-25} \ \text{ecm}.
 \label{cedm}
 \end{equation}
 This bound severely constrains phases of the mixing parameters $\delta^{dL(dR)}_{23}$ 
of Eq.(\ref{mixing}).


\section{Tension between $\epsilon_K$ and $\sin 2\beta$}
\label{sec:tension}

We start our numerical discussion by looking at the $\epsilon_K$ parameter,
which is given in the following theoretical formula
\begin{equation}
\epsilon_K
=
e^{i \phi_{\epsilon}} \sin{\phi_{\epsilon}} \left( \frac{\text{Im}(M_{12}^K)}{\Delta M_K}
+ \xi \right),  \qquad
\xi
=
\frac{\text{Im} A_0}{\text{Re} A_0} , \qquad
\phi_\epsilon=\tan^{-1}\left( \frac{2 \Delta M_K}{\Delta \Gamma_K} \right),
\end{equation}
with $A_0$ being the isospin zero amplitude in $K\to\pi\pi$ decays.
Here, $M_{12}^K$ is the dispersive part of the  $K^0-\bar{K^0}$ mixing, $\Delta M_K$ is the mass difference in the neutral K meson.
An effect of suppression factor $\kappa_{\epsilon}$ which indicates effects of $\xi \ne 0$ and $\phi_{\epsilon} < \pi/4$,
 was given by Buras and Guadagnoli \cite{Buras:2008nn} as:
\begin{equation}
\kappa_{\epsilon}
=
0.92 \pm 0.02 \ \ .
\end{equation}

In the SM, the dispersive part $M_{12}^K$ is given as follows, 
\begin{align}
M_K^{12}
&=
\langle K| \mathcal{H}_{\Delta F=2} |\bar{K} \rangle \nonumber \\
&=
-\frac{4}{3}\left( \frac{G_F}{4 \pi} \right)^2 M_W^2 \hat{B}_K F_K^2 M_K \left( \eta_{cc} \lambda_c^2 E(x_c)
+\eta_{tt} \lambda_t^2 E(x_t)
+2 \eta_{ct} \lambda_c \lambda_t E(x_c,x_t) \right) ,
\end{align}
where $\lambda_c = V_{cs}V_{cd}^*,\  \lambda_t = V_{ts}V_{td}^*$,
and $E(x)$'s are the one-loop functions \cite{Inami:1980fz}.
Then, we obtain $|\epsilon_K^{\text{SM}}|$ in terms of the Wolfenstein parameters
 $\lambda$, $\rho$ and $\eta$ as follows:
\begin{align}
|\epsilon_K^{\text{SM}}|
&=
\kappa_{\epsilon} C_{\epsilon} \hat{B}_K |V_{cb}|^2 \lambda^2 \bar{\eta} 
\left( |V_{cb}|^2 (1-\bar{\rho})\eta_{tt} E(x_t)
-\eta_{cc}E(x_c)
+\eta_{ct} E(x_c,x_t) \right) \nonumber \\
&=
\kappa_{\epsilon} C_{\epsilon} \hat{B}_K |V_{cb}|^2 \lambda^2
\left(
\frac{1}{2} |V_{cb}|^2 R_t^2 \sin(2 \beta) \eta_{tt} E(x_t)
+R_t \sin \beta (-\eta_{cc}E(x_c) +\eta_{ct} E(x_c,x_t))
\right),
\label{epsilonKSM}
\end{align}
where 
\begin{align}
 C_{\epsilon}=
\frac{G_F^2 F_K^2 m_K M_W^2}{6  \sqrt{2} \pi^2 \Delta M_K} ,
\end{align}
and
\begin{align}
\bar\rho=\rho \  \left (1-\frac{1}{2}\lambda^2\right ),
\qquad \bar\eta=\eta \  \left (1-\frac{1}{2}\lambda^2 \right ).
\end{align}
In Eq.(\ref{epsilonKSM}), we use the relation: 
\begin{align}
R_t \sin{\beta}
= \bar{\eta}, \qquad
R_t \cos{\beta}
= 1- \bar{\rho},
\end{align}
where $R_t$ is
\begin{align}
R_t
=
\frac{1}{\lambda}
\frac{|V_{td}|}{|V_{ts}|} 
=
\frac{1}{\lambda}
\frac{F_{B_s}\sqrt{B_s}}{F_{B}\sqrt{B}}
\sqrt{\frac{M_{{B_s}}}{M_{B^0}}}
\sqrt{\frac{\Delta M_{B^0}^{\rm exp}}{\Delta M_{B_s}^{\rm exp}}}.
\end{align}
As seen in Eq.(\ref{epsilonKSM}),  $|\epsilon_K^{\text{SM}}|$ is given 
in terms of  $\sin(2\beta)$  because
 there is only one CP violating phase in the SM.

 If we take into account the  gluino-sbottom-quark interaction, 
$\epsilon_K$ is modified as
 \begin{align}
  \epsilon_K=\epsilon_K^{\text{SM}}+\epsilon_K^{\tilde g}.
  \label{epsilon}
  \end{align}
Here, $\epsilon_K^{\tilde g}$ is given by the imaginary part of the gluino-sbottom box diagram, which is presented in  Appendix A. 
 The magnitude of $\epsilon_K^{\tilde g}$ is proportional to 
 the product $|\delta _{13}^{dL(dR)}\times \delta _{23}^{dL(dR)}|$ because the first and second families are decoupled in the  gluino-sbottom box diagrams.
 We should also modify $R_t$ as follows:
 \begin{align}
R_t
=
\frac{1}{\lambda}
\frac{F_{B_s}\sqrt{B_s}}{F_{B}\sqrt{B}}
\sqrt{\frac{M_{B_s}}{M_B}}
\sqrt{\frac{\Delta M_{B^0}^{\rm exp}}{\Delta M_{B_s}^{\rm exp}}}
\sqrt{\frac{C_s}{C_d}},
\end{align}
where
\begin{align}
C_q= 1+h_q e^{2i\sigma _q}, \quad (q=d, \ s).
\end{align}

Now, we remark the non-perturbative parameter $\hat{B}_K$ in eq.(\ref{epsilonKSM}).  
Recently, the error of this parameter shrank dramatically in the lattice calculations.
The most updated value is presented as \cite{Ciuchini-KEKFF2013,Flynn-KEKFF2013}
 \begin{equation}
\hat B_K=
0.73 \pm 0.03 \ \ .
\label{BK}
\end{equation}
By inputting this value, we can calculate $|\epsilon_K^{\text{SM}}|$
 for the fixed  $\sin(2\beta)$.
In other words, we can test numerically  the overlap region of
 among $|\epsilon_K|$, ${\Delta M_{B^0}}/{\Delta M_{B_s}}$
 and $\sin(2\beta)$ in the unitarity triangle of the SM.
\vskip 0 cm
\begin{wrapfigure}{l}{8.5cm}
\begin{center}
\includegraphics[width=7.3cm]{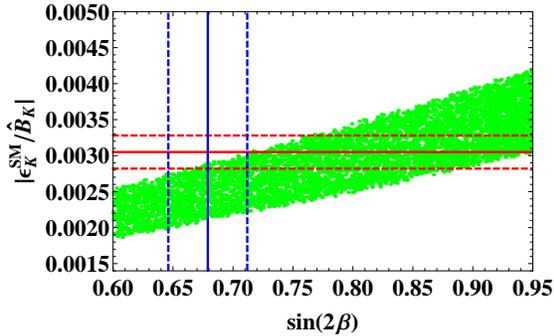}
\caption{The predicted region on $\sin 2\beta$-$|\epsilon_K|/\hat B_K$ plane in SM. 
Solid and dotted lines denote the experimental  best fit 
and bounds  with $90\% $~C.L.}
\label{tension}
\end{center}
 \end{wrapfigure}

We obtain the relation between   $\sin(2\beta)$ and $|\epsilon_K^{\text{SM}}/\hat B_K|$, which is shown
  with the experimental allowed region with $90\%$ C.L. in Figure \ref{tension}.  
 It is noticed that the consistency between the SM prediction and the experimental data 
in  $\sin(2\beta)$ and  $|\epsilon_K^{\text{SM}}/\hat B_K|$ is marginal.
 This fact was pointed out by Buras and Guadagnoli \cite{Buras:2008nn}, and called as the tension between $|\epsilon_K|$ and $\sin(2\beta)$.
 This situation may indicate the new physics.
We will show that this tension is understood by taking account of
 the SUSY box diagram through the gluino-sbottom-quark interaction, which also predicts the deviation
 from the SM in  the CP violations of
$B^0\to \phi K_S$, $B^0\to \eta ' K^0$,  $B_s\to \phi\phi$,  $B_s\to \phi \eta'$,
  $B^0 \to K^0 \bar K^0$,  $B^0\to \mu ^-X$, and  $B_s\to \mu ^-X$ decays.

\section{Numerical Results}

 Let us present the numerical results.
 In order to constrain the gluino-sbottom-quark mixing parameters,
 we input the experimental data of the CP violations,
 $\epsilon_K$, $\phi_d$, and $\phi_s$.
 The experimental upper bound of cEDM of the strange quark is also put.
 In addition to these experimental data of the CP violations and T violation, we take account of the
 observed values $\Delta M_{B^0}$, $\Delta M_{B_s}$, the CKM mixing $|V_{ub}|$,
 and the branching ratio of $b\to s\gamma$.
%
\begin{figure}[h]
\begin{minipage}[]{0.45\linewidth}
\includegraphics[width=7.8cm]{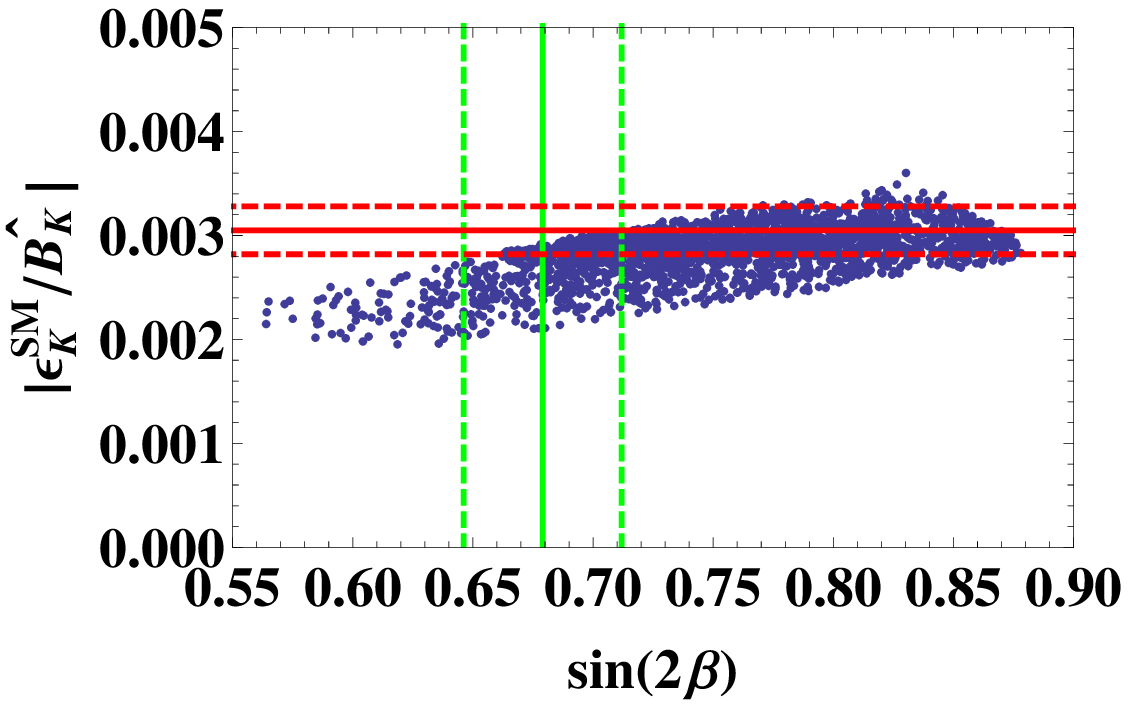}

\hspace{1cm}
\caption{Predicted region on $|\epsilon_K^{\text{SM}}|/\hat B_K$-$\sin(2\beta)$ plane. 
Vertical and horizontal dashed lines denote the experimental allowed region with $90\%$C.L. Vertical and horizontal solid lines denote
observed  central values.} 
\label{tensionSUSY}
\end{minipage}
\hspace{1cm}
\begin{minipage}[]{0.45\linewidth}
\includegraphics[width=7.5cm]{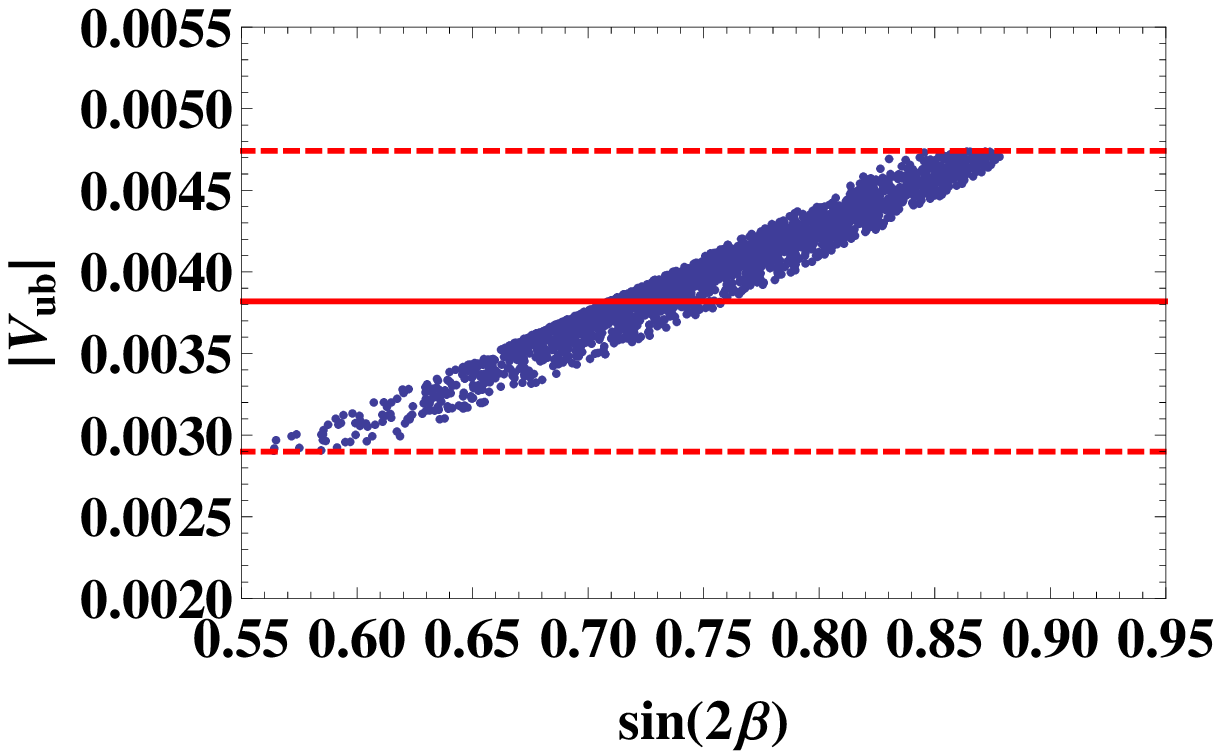}
\hspace{1cm}
\caption{The predicted $|V_{ub}|$ versus  $\sin(2\beta)$. Horizontal dashed lines denote the experimental allowed region with $90\%$C.L. of $|V_{ub}|$. Horizontal solid lines denote the observed  central values.} 
\label{Vub}
\end{minipage}
\end{figure}
The input parameters in our calculation  are summarized in Table 1.

\begin{table}[t]
\begin{center}
\begin{tabular}{|l|}
\hline
$\alpha_s(M_Z)=0.1184$ \cite{PDG}\\
$m_c(m_c)=1.275$ GeV \cite{PDG}\\
$m_t(m_c)=1.275$ GeV $(\bar{MS})$ \cite{PDG}\\
$M_{B_s}=5.36677(24)$ GeV \cite{PDG}\\
$\Delta M_s=(116.942 \pm 0.1564)\times 10^{-13}$ GeV \cite{Aaij:2013mpa}\\
$\Delta M_d=(3.337 \pm 0.033)\times 10^{-13}$ GeV \cite{PDG} \\
$f_{B_s} = (233\pm 10)$ MeV \cite{Ciuchini-KEKFF2013}\\
$f_{B_s}/f_{B^0} = 1.200\pm 0.02$ \cite{Ciuchini-KEKFF2013}\\
$\xi_s=1.21(6)$ \cite{Buras:2008nn}\\
$\lambda=0.2255(7)$ \cite{PDG}\\
$|V_{cb}|=(4.12\pm0.11)\times 10^{-2}$  \cite{Ciuchini-KEKFF2013}\\
$\eta_{cc}=1.43(23)$ \cite{Buras:2008nn}\\
$\eta_{ct}=0.47(4)$ \cite{Buras:2008nn}\\
$\eta_{tt}=0.5765(65)$ \cite{Buras:2008nn}\\
$f_K = (156.1\pm1.1)$ MeV \cite{PDG}\\ 
$\kappa_{\epsilon}=0.92(2)$ \cite{Buras:2008nn}\\
\hline
\end{tabular}
\end{center}
\caption{Input parameters in our calculation.}
\label{tab:inputparameters}
\end{table}

 At first, we present the allowed region on the plane of
  $|\epsilon_K^{\text{SM}}|/\hat B_K$ and $\sin(2\beta)$ in Figure \ref{tensionSUSY},
where  SM components in Eqs. (\ref{phiSUSY}) and (\ref{epsilon})  are only
shown. 
 The present experimental data of $\sin\phi_d$ in Eq. (\ref{phasedata}) 
 allows the range of   $\sin(2\beta)=0.57-0.88$, where $\beta$ is one angle of the unitarity triangle.
 Once we take account of the contribution of the gluino-sbottom-quark
interaction,  the allowed regions of   $|\epsilon_K|$ and $\sin\phi_d$ converge 
 within the experimental error-bars.
 
 When $\sin 2\beta$ and $\Delta M_{B^0}^{\rm SM}/\Delta M_{B_s}^{\rm SM}$ are fixed,
 the  $|V_{ub}|$ is predicted.
 In Figure \ref{Vub}, we show the relation between  $\sin(2\beta)$ and $|V_{ub}|$, 
 where the outside  the experimental error-bar of  $|V_{ub}|$ are cut.
 In the present measurement of $|V_{ub}|$, there is $2.6~\sigma$ discrepancy
 in the exclusive and inclusive decays as follows \cite{Ciuchini-KEKFF2013}:
  \begin{equation}
  |V_{ub}|=(3.28\pm 0.30) \times 10^{-3} \ (\text{exclusive}), \quad
   \quad
 |V_{ub}|=(4.40\pm 0.31) \times 10^{-3} \ (\text{inclusive}),
  \end{equation}
  although the average value is $|V_{ub}|=(3.82\pm 0.56) \times 10^{-3}$.
  The precise observation of  $|V_{ub}|$ leads to the 
determination of $\sin (2 \beta)$.

%

\begin{figure}[h!]
\begin{minipage}[]{0.45\linewidth}
\hspace{4cm}
\includegraphics[width=7.5cm]{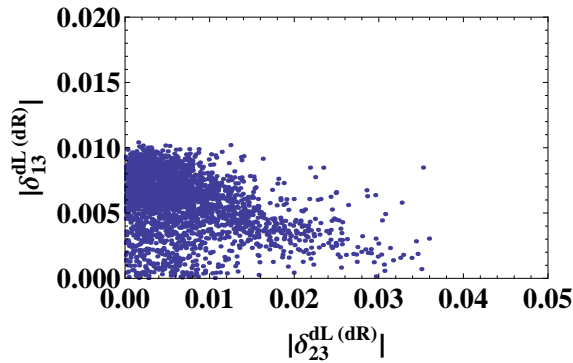}
\end{minipage}
\caption{Allowed region of the mixing parameters, $|\delta_{13}^{dL(dR)}|$ and
 $|\delta_{23}^{dL(dR)}|$. } 
\label{SUSYparameters}
\hspace{1cm}
\end{figure}

 
The allowed region of the mixing parameters  $|\delta_{13}^{dL(dR)}|$ and 
$|\delta_{23}^{dL(dR)}|$ are shown in Figure \ref{SUSYparameters},
where
we input the experimental data of the CP violations,
 $\epsilon_K$, $\phi_d$, and $\phi_s$.
 The experimental upper bound of cEDM of the strange quark is also input.
 We also take account of the
 observed values $\Delta M_{B^0}$, $\Delta M_{B_s}$, the CKM mixing $|V_{ub}|$,
 and the branching ratio of $b\to s\gamma$.

 As seen in Figure \ref{SUSYparameters}, we obtain the allowed region of 
\begin{equation}
|\delta_{13}^{dL(dR)}|=0\sim 0.01, \qquad   |\delta_{23}^{dL(dR)}|=0\sim 0.04.
\end{equation}
By using these values, we discuss the sensitivity of the SUSY contribution 
to the CP violation of the $B^0$ and $B_s$ decays.

\begin{figure}[h!]
\begin{minipage}[]{0.45\linewidth}
\includegraphics[width=7.5cm]{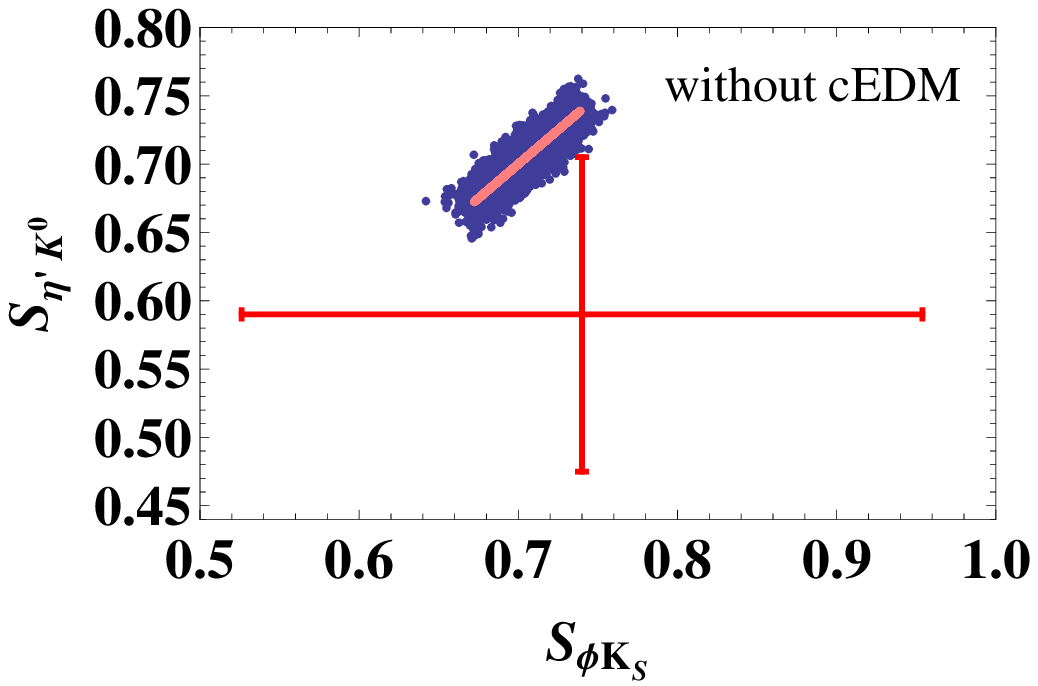}
\end{minipage}
\hspace{1cm}
\begin{minipage}[]{0.45\linewidth}
\includegraphics[width=7.5cm]{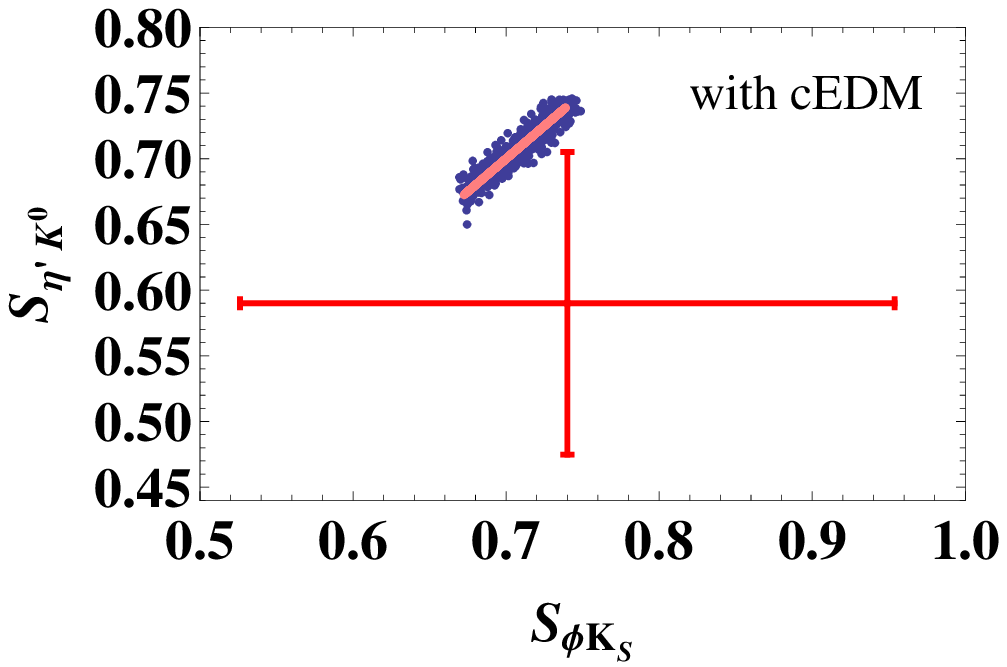}
\end{minipage}
\caption{The predicted time dependent CP asymmetries 
on  $S_{\phi K_S}$--$S_{\eta 'K^0}$ plane without/with the constraint
of  cEDM of the strange quark.
The SM prediction $S_{J/\psi K_S}\simeq S_{\phi K_S}=S_{\eta 'K^0}$ 
is plotted by the pink slant lines. 
The experimental data with error-bar is plotted by the red solid lines at $90\%$ C.L.}
\label{SphiKSetapK}
\end{figure}

Let us discuss the time dependent CP asymmetries 
$S_{\phi K_S}$ and $S_{\eta 'K^0}$.
The SM leads to  $S_{J/\psi K_S}\text{(SM)}\simeq S_{\phi K_S}\text{(SM)}=
S_{\eta 'K^0}\text{(SM)}$, 
while the present data of these time dependent CP asymmetries are given in Eq.~(\ref{Sfdata}). 
We predict the deviation from the SM in  Figure~\ref{SphiKSetapK} for the two cases, 
where the constraint of the cEDM of the strange quark is imposed or is not imposed.
It is clearly seen that the cEDM of the strange quark reduces the deviation 
from the SM.
Thus, it is very difficult to observe the gluino-sbottom-quark contribution in these
non-leptonic decays.

\begin{figure}[h!]
\begin{minipage}[]{0.45\linewidth}
\includegraphics[width=7.5cm]{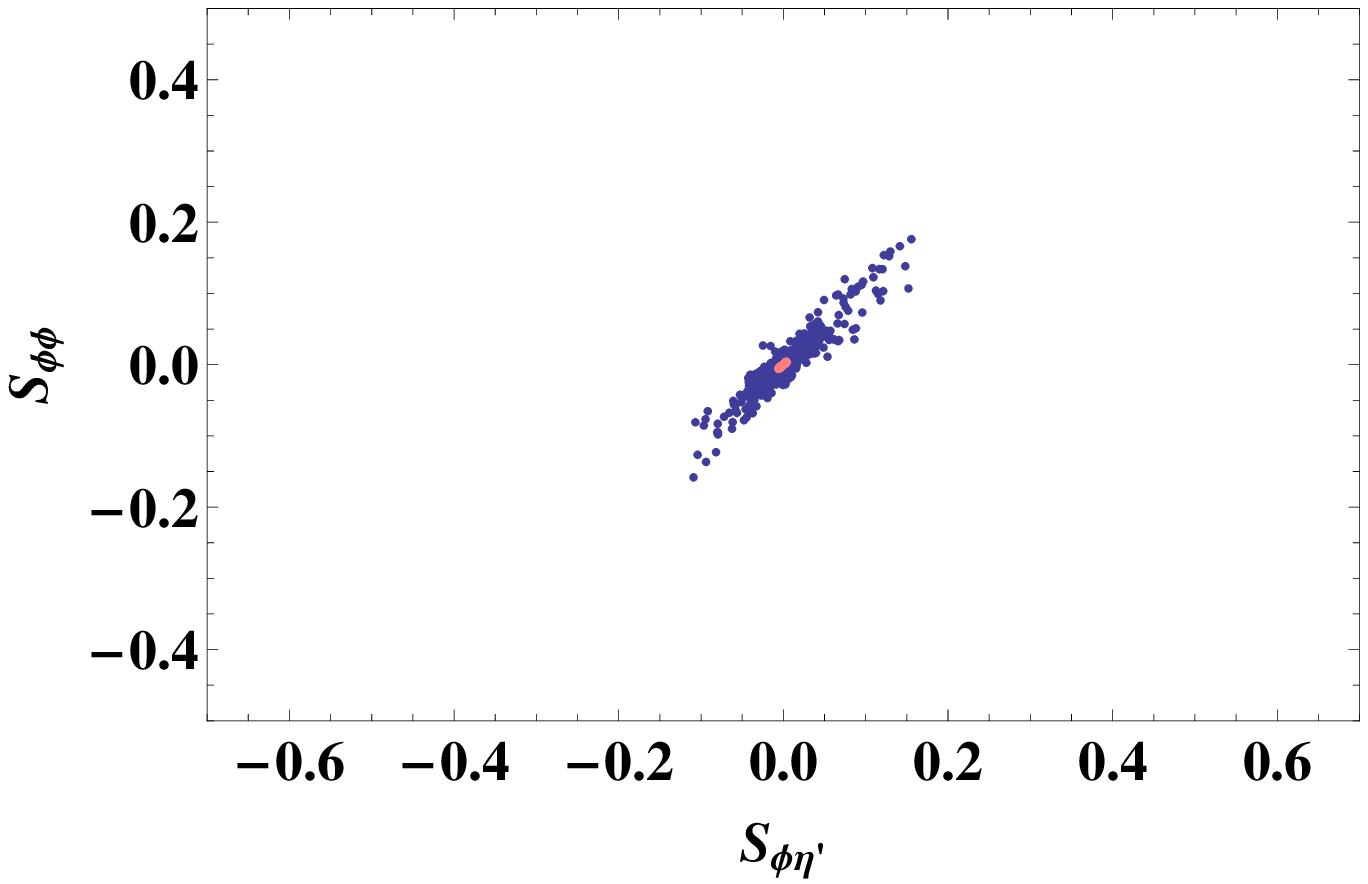}
\caption{The predicted time dependent CP asymmetries on 
 $S_{\phi \eta '}$--$S_{\phi\phi }$ plane. 
The small pink region  denotes the SM prediction.} 
\label{fig:SphiKSetapK}
\end{minipage}
\hspace{1cm}
\begin{minipage}[]{0.45\linewidth}
\includegraphics[width=7.5cm]{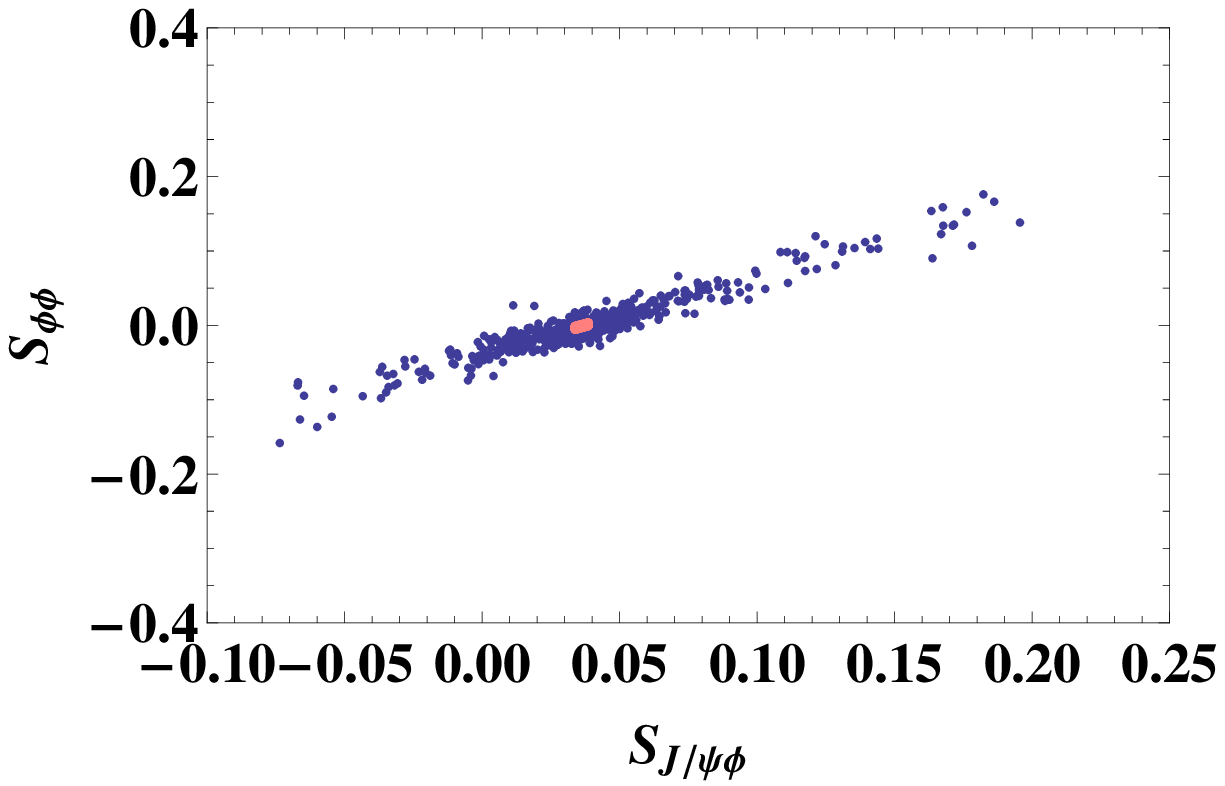}
\caption{The predicted  
 $S_{\phi\phi }$ versus $S_{J/\psi\phi}$ plane, where
  $S_{J/\psi\phi}$ is plotted within the experimental error at  $90\%$ C.L.
The small pink region denotes the SM prediction.
} 
\label{fig:SJpsiphiSphiphi}
\end{minipage}
\end{figure}

In Figure ~\ref{fig:SphiKSetapK}, we show the prediction of  the time dependent CP asymmetries $S_{\phi \phi }$ and $S_{\phi \eta '}$, where the constraint of
the cEDM is imposed. 
We use the experimental result of $S_{J/\psi \phi }$ for 
the phase $\phi_s$, which is given in Eq.~(\ref{phasedata}), in our calculations. 
We denote the small pink region as the SM value 
$S_{J/\psi \phi }(\text{SM})=-0.0363\pm 0.0017$~\cite{Charles:2004jd} in the figure. 
It is found that  $S_{J/\psi \phi }$ is almost proportional to $S_{\phi \phi}$.
If the $\Delta B=1$ SUSY contribution is seizable, these asymmetries should be  different
each other as seen in Eq.(\ref{phi-eta}).
That is, the gluino interaction induced  $\Delta B=1$ contribution
is very small.  On the other hand, the gluino
 induced  $\Delta B=2$ contribution (SUSY box diagrams) could be detectable
as seen in Eqs.(\ref{phiSUSY}) and (\ref{asymBs}). 
 This situation is understandable because the magnitude of 
 $C_{8G}^{\tilde g}(m_b)$ is reduced
by the cancellation between the contributions of two sbottom
 $\tilde b_1$ and $\tilde b_2$ as seen in Appendix B.
In conclusion, we predict $-0.1\lesssim S_{\phi \phi }\lesssim 0.2$ 
and $-0.1\lesssim S_{\phi \eta '}\lesssim 0.2$, respectively. 
Since the phase $\phi _s$ has still large experimental error bar, 
our prediction will be improved if the precise experimental data of 
$S_{J/\psi \phi }$ will be given in the near future at LHCb. 
In order to see this situation clearly,
we show the $S_{J/\psi\phi}$ dependence for the predicted $S_{\phi \phi }$ 
 in  Figure \ref{fig:SJpsiphiSphiphi}.
 
LHCb reported the first
flavor-tagged measurement of the time-dependent CP-violating asymmetry
in $B_s\to \phi\phi$  decay \cite{Aaij:2013qha}.
The CP-violating phase is measured to be in the interval $\phi_s=[-2.46,-0.76]$ rad
 as seen in Eq.(\ref{phiphidata}).
The precision of the CP violating phase measurement is dominated by the statistical uncertainty and is expected to improve with larger LHCb data in the near future.

\begin{figure}[h!]
\begin{minipage}[]{0.45\linewidth}
\includegraphics[width=7.5cm]{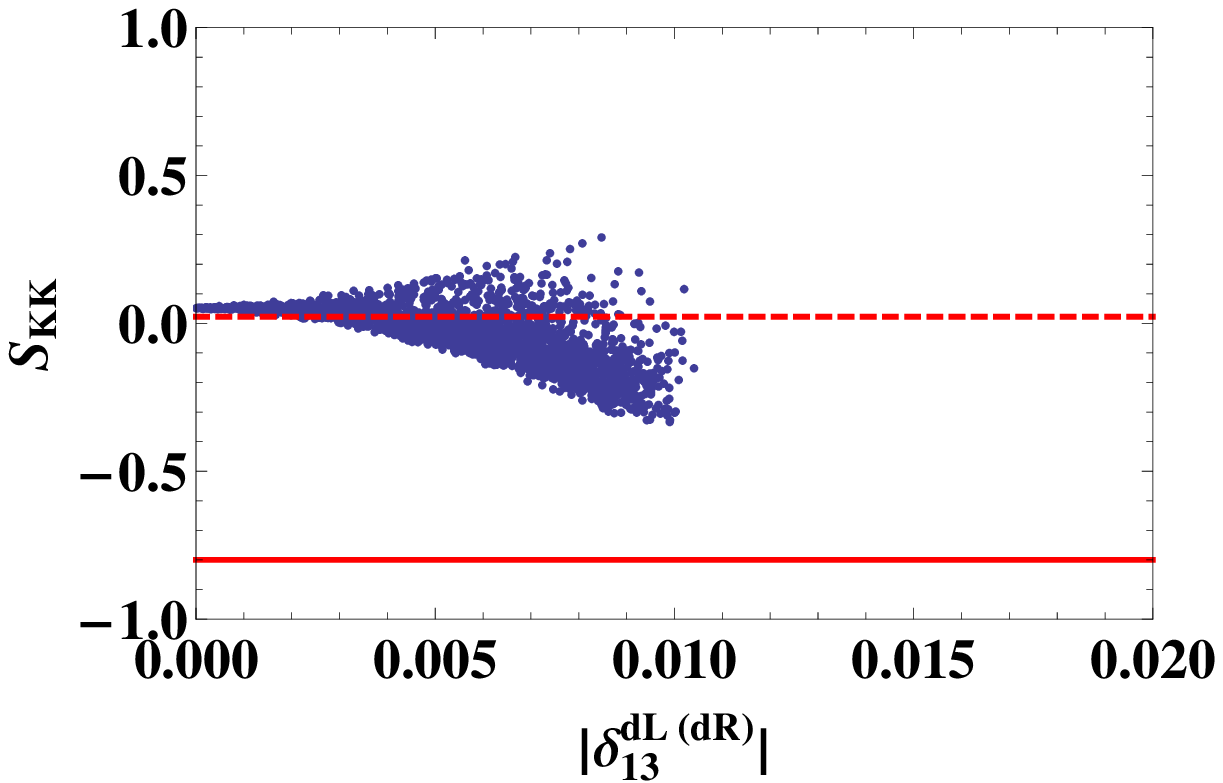}
\caption{The predicted time dependent CP asymmetry $S_{K^0\bar K^0}$ 
versus $|\delta^{dL(dR)}_{13}|$. 
The red solid and red dotted lines denote the best fit value 
and the experimental bound with $90\% $~C.L., respectively.} 
\label{fig:MISKK}
\end{minipage}
\hspace{1cm}
\begin{minipage}[]{0.45\linewidth}
\includegraphics[width=7.5cm]{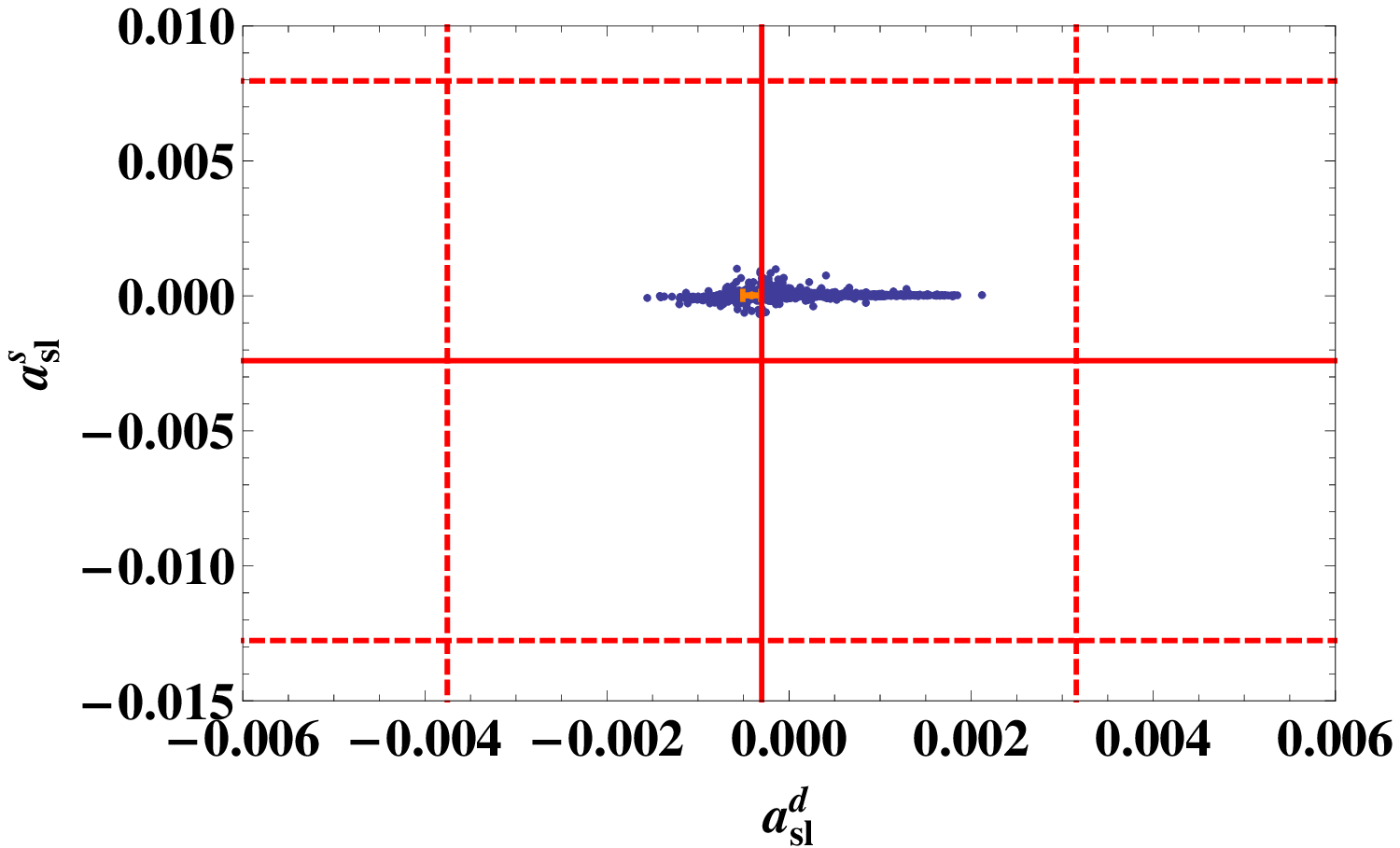}
\caption{Predicted semi-leptonic CP asymmetries $a_{sl}^d$ and $a_{sl}^s$.
The red solid and red dotted lines denote the best fit value 
and the experimental bounds  with $90\% $~C.L., respectively.} 
\label{fig:semi}
\end{minipage}
\end{figure}

In Figure~\ref{fig:MISKK} , 
we show the prediction of the time dependent CP asymmetry
$S_{K^0\bar K^0}$  depending on $|\delta^{dL(dR)}_{13}|$.
The predicted region  is  $-0.4 \le S_{K^0\bar K^0}\le 0.3$, on the other hand,
one predicts $S_{K^0\bar K^0}(\text{SM}) \simeq 0.06$ in the SM
~\cite{Giri:2004af}.
The present experimental data are given as $S_{K^0\bar K^0}(\text{exp})=-0.8\pm 0.5$ 
\cite{PDG}. 
Since the SM predicted value is tiny, we have a chance to observe the  SUSY contribution
by the  precise experimental data in the near future.

At last, we present   the prediction of the indirect CP violation 
 $a_{sl}^d$ and $a_{sl}^s$
in Figure \ref{fig:semi}.  The predicted region is 
$a_{sl}^d=-0.0017 \sim 0.002$ and $a_{sl}^s=-0.001 \sim 0.001$,
on the other hand, the SM gives
   $a^{d\rm SM}_{sl}=-(4.1\pm 0.6)\times 10^{-4}$ and 
$a^{s\rm SM}_{sl}=(1.9\pm 0.3)\times 10^{-5}$
as shown in Eq.(\ref{semiCP}).
The experimental data still have large error-bars \cite{Vesterinen:2013jia,PDG}.
 The precise measurement of the semi-leptonic asymmetry $a_{sl}^d$ at Belle II will provide us
 an interesting test of the SUSY contribution.

\section{Summary}

We have discussed the sensitivity of the gluino-sbottom-quark interaction
 to the CP violating phenomena of the  $K$, $B^0$ and $B_s$ mesons.
We take the split-family scenario, which is the consistent with the  LHC data. In this scenario,  the first and second family
 squarks are very heavy, ${\cal O}(10)$~TeV, on the other hand, the third family
  squark masses are  at  ${\cal O}(1)$~TeV.
  Then,  the  $s\to d$ transition is mediated by the second order contribution of  the third family sbottom.
  We have used $m_{\tilde g}=2$~TeV, $m_{\tilde b_1}=1$~TeV, and  $m_{\tilde b_2}=1.1$~TeV.
  In order to constrain the gluino-sbottom-quark mixing parameters,
 we input the experimental data of the CP violations,
 $\epsilon_K$, $\phi_d$, and  $\phi_s$.
 The experimental upper bound of the cEDM of the strange quark is also input.
 In addition,  we take account of the observed values
 $\Delta M_{B^0}$, $\Delta M_{B_s}$, the CKM mixing $|V_{ub}|$,
 and the branching ratio of $b\to s\gamma$.

 By using the non-perturbative parameter $\hat{B}_K=0.73 \pm 0.03$, which is the most updated value in the lattice calculations, it is clearly presented that
the consistency between the SM prediction and the experimental data 
is marginal on the $\sin(2\beta)-|\epsilon_K^{\text{SM}}|$ plane.
  This tension has been solved by taking account of the gluino-sbottom-quark interaction.

 The allowed region of the mixing parameters   are obtained as
$|\delta_{13}^{dL(dR)}|=0\sim 0.01$ and $|\delta_{23}^{dL(dR)}|=0\sim 0.04$.
By using these values, the deviations from the SM prediction are estimated 
in the CP violation of  $B^0$ and $B_s$ decays.
The CP asymmetries of the 
 $B^0\to \phi K_S$ and $B^0\to \eta 'K^0$ decays are found to be tiny due to the  cEDM constraint of the strange quark.

On the other hand,  the CP asymmetries  of the $B_s\to \phi \phi $ 
and $B_s\to \phi \eta '$ decays  could be largely deviated from the SM predictions 
such as
 $-0.1\lesssim S_{\phi \phi }\lesssim 0.2$ and $-0.1\lesssim S_{\phi \eta '}\lesssim 0.2$.
It is remarked  that  the time dependent CP asymmetry $S_{\phi \phi }$  is almost proportional to $S_{\phi \eta '}$.
That is, the gluino-sbottom interaction induced  $\Delta B=1$ transition
is very small, but the $\Delta B=2$ transition  could be detectable. 
Since the phase $\phi _s$ has still large experimental error-bar, 
our prediction will be improved if the precise experimental data of 
$S_{J/\psi \phi }$ will be given in the near future at the LHCb experiment. 

We also  predict the time dependent CP asymmetry 
$S_{K^0\bar K^0}$  as $-0.4 \le S_{K^0\bar K^0}\le 0.3$ while
one predicts $S_{K^0\bar K^0}(\text{SM})\simeq 0.06$ in the SM. 
 More precise data will test the  SUSY contribution in the near future.
The semi-leptonic CP asymmetries 
 $a_{sl}^d$ and $a_{sl}^s$  are predicted as
$a_{sl}^d=-0.0017 \sim 0.002$ and $a_{sl}^s=-0.001 \sim 0.001$,
while the SM predicts
   $a^{d\rm SM}_{sl}=-(4.1\pm 0.6)\times 10^{-4}$ and 
$a^{s\rm SM}_{sl}=(1.9\pm 0.3)\times 10^{-5}$. 
 We expect the precise measurement of the  $a_{sl}^d$ at Belle II, which  will provide us
  interesting tests of the squark flavor mixing.

\vspace{0.5 cm}
\noindent
{\bf Acknowledgment}

This work is   supported by JSPS Grand-in-Aid for Scientific Research,
 21340055 and 24654062, 25-5222, respectively.
\appendix{}
\section*{Appendix}
\section{Squark contribution in $\Delta F=2$ process}

The $\Delta F=2$ effective Lagrangian from the gluino-sbottom-quark interaction  is given as
\begin{align}
\mathcal{L}_{\text{eff}}^{\Delta F=2}=&-\frac{1}{2}\left [C_{VLL}O_{VLL}+C_{VRR}O_{VRR}\right ] \nonumber \\
&-\frac{1}{2}\sum _{i=1}^2
\left [C_{SLL}^{(i)}O_{SLL}^{(i)}+C_{SRR}^{(i)}O_{SRR}^{(i)}+C_{SLR}^{(i)}O_{SLR}^{(i)}\right ],
\label{Lagrangian-DeltaF=2}
\end{align}
then, the $P^0$-$\bar P^0$ mixing, $M_{12}$, is written as 
\begin{equation}
M_{12}=-\frac{1}{2m_P}\langle P^0|\mathcal{L}_{\text{eff}}^{\Delta F=2}|\bar P^0\rangle \ .
\end{equation}
The hadronic matrix elements are given in terms of the non-perturbative
parameters  $B_i$ as: 
\begin{align}
\langle P^0|\mathcal{O}_{VLL}|\bar P^0\rangle &=\frac{2}{3}m_P^2f_P^2B_1, \quad 
\langle P^0|\mathcal{O}_{VRR}|\bar P^0\rangle =\langle P^0|\mathcal{O}_{VLL}|\bar P^0\rangle ,\nonumber \\
\langle P^0|\mathcal{O}_{SLL}^{(1)}|\bar P^0\rangle &=-\frac{5}{12}m_P^2f_P^2R_PB_2, \quad 
\langle P^0|\mathcal{O}_{SRR}^{(1)}|\bar P^0\rangle =\langle P^0|\mathcal{O}_{SLL}^{(1)}|\bar P^0\rangle ,\nonumber \\
\langle P^0|\mathcal{O}_{SLL}^{(2)}|\bar P^0\rangle &=\frac{1}{12}m_P^2f_P^2R_PB_3, \quad 
\langle P^0|\mathcal{O}_{SRR}^{(2)}|\bar P^0\rangle =\langle P^0|\mathcal{O}_{SLL}^{(2)}|\bar P^0\rangle ,\nonumber \\
\langle P^0|\mathcal{O}_{SLR}^{(1)}|\bar P^0\rangle &=\frac{1}{2}m_P^2f_P^2R_PB_4, \quad 
\langle P^0|\mathcal{O}_{SLR}^{(2)}|\bar P^0\rangle =\frac{1}{6}m_P^2f_P^2R_PB_5,
\end{align}
where 
\begin{equation}
R_P=\left (\frac{m_P}{m_Q+m_q}\right )^2,
\end{equation}
with $(P,Q,q)=(B_d,b,d),~(B_s,b,s),~(K,s,d)$.

The Wilson coefficients for the gluino contribution in Eq.~(\ref{Lagrangian-DeltaF=2}) are written as \cite{GotoNote}

\begin{align}
C_{VLL}(m_{\tilde g})&=\frac{\alpha _s^2}{m_{\tilde g}^2}\sum _{I,J=1}^6
(\lambda _{GLL}^{(d)})_I^{ij}(\lambda _{GLL}^{(d)})_J^{ij}
\left [\frac{11}{18}g_{2[1]}(x_I^{\tilde g},x_J^{\tilde g})
+\frac{2}{9}g_{1[1]}(x_I^{\tilde g},x_J^{\tilde g})\right ],\nonumber \\
C_{VRR}(m_{\tilde g})&=C_{VLL}(m_{\tilde g})(L\leftrightarrow R),\nonumber \\
C_{SRR}^{(1)}(m_{\tilde g})&=\frac{\alpha _s^2}{m_{\tilde g}^2}\sum _{I,J=1}^6
(\lambda _{GLR}^{(d)})_I^{ij}(\lambda _{GLR}^{(d)})_J^{ij}
\frac{17}{9}g_{1[1]}(x_I^{\tilde g},x_J^{\tilde g}),\nonumber \\
C_{SLL}^{(1)}(m_{\tilde g})&=C_{SRR}^{(1)}(m_{\tilde g})(L\leftrightarrow R),\nonumber \\
C_{SRR}^{(2)}(m_{\tilde g})&=\frac{\alpha _s^2}{m_{\tilde g}^2}\sum _{I,J=1}^6
(\lambda _{GLR}^{(d)})_I^{ij}(\lambda _{GLR}^{(d)})_J^{ij}
\left (-\frac{1}{3}\right )g_{1[1]}(x_I^{\tilde g},x_J^{\tilde g}),\nonumber \\
C_{SLL}^{(2)}(m_{\tilde g})&=C_{SRR}^{(2)}(m_{\tilde g})(L\leftrightarrow R),\nonumber \\
C_{SLR}^{(1)}(m_{\tilde g})&=\frac{\alpha _s^2}{m_{\tilde g}^2}\sum _{I,J=1}^6
\Bigg \{ (\lambda _{GLR}^{(d)})_I^{ij}(\lambda _{GRL}^{(d)})_J^{ij}
\left (-\frac{11}{9}\right )g_{2[1]}(x_I^{\tilde g},x_J^{\tilde g}) \nonumber \\
&\hspace{2cm}+(\lambda _{GLL}^{(d)})_I^{ij}(\lambda _{GRR}^{(d)})_J^{ij}
\left [\frac{14}{3}g_{1[1]}(x_I^{\tilde g},x_J^{\tilde g})
-\frac{2}{3}g_{2[1]}(x_I^{\tilde g},x_J^{\tilde g})\right ]\Bigg \} ,\nonumber \\
C_{SLR}^{(2)}(m_{\tilde g})&=\frac{\alpha _s^2}{m_{\tilde g}^2}\sum _{I,J=1}^6
\Bigg \{ (\lambda _{GLR}^{(d)})_I^{ij}(\lambda _{GRL}^{(d)})_J^{ij}
\left (-\frac{5}{3}\right )g_{2[1]}(x_I^{\tilde g},x_J^{\tilde g}) \nonumber \\
&\hspace{2cm}+(\lambda _{GLL}^{(d)})_I^{ij}(\lambda _{GRR}^{(d)})_J^{ij}
\left [\frac{2}{9}g_{1[1]}(x_I^{\tilde g},x_J^{\tilde g})
+\frac{10}{9}g_{2[1]}(x_I^{\tilde g},x_J^{\tilde g})\right ]\Bigg \} ,
\end{align}
where
\begin{align}
(\lambda _{GLL}^{(d)})_K^{ij}&=(\Gamma _{GL}^{(d)\dagger })_i^K(\Gamma _{GL}^{(d)})_K^j~,\quad 
(\lambda _{GRR}^{(d)})_K^{ij}=(\Gamma _{GR}^{(d)\dagger })_i^K(\Gamma _{GR}^{(d)})_K^j~,\nonumber \\
(\lambda _{GLR}^{(d)})_K^{ij}&=(\Gamma _{GL}^{(d)\dagger })_i^K(\Gamma _{GR}^{(d)})_K^j~,\quad 
(\lambda _{GRL}^{(d)})_K^{ij}=(\Gamma _{GR}^{(d)\dagger })_i^K(\Gamma _{GL}^{(d)})_K^j~.
\end{align}
Here we take $(i,j)=(1,3),~(2,3),~(1,2)$ which correspond to $B^0$, $B_s$, and $K^0$ mesons, respectively. 
The loop functions are given as follows:
\begin{itemize}
\item If $x_I^{\tilde g}\not =x_J^{\tilde g}$ ($x_{I,J}^{\tilde g}=m_{\tilde d_{I,J}}^2/m_{\tilde g}^2$),
\begin{align}
g_{1[1]}(x_I^{\tilde g},x_J^{\tilde g})&=\frac{1}{x_I^{\tilde g}-x_J^{\tilde g}}
\left (\frac{x_I^{\tilde g}\log x_I^{\tilde g}}{(x_I^{\tilde g}-1)^2}
-\frac{1}{x_I^{\tilde g}-1}-\frac{x_J^{\tilde g}\log x_J^{\tilde g}}{(x_J^{\tilde g}-1)^2}
+\frac{1}{x_J^{\tilde g}-1}\right ),\nonumber \\
g_{2[1]}(x_I^{\tilde g},x_J^{\tilde g})&=\frac{1}{x_I^{\tilde g}-x_J^{\tilde g}}
\left (\frac{(x_I^{\tilde g})^2\log x_I^{\tilde g}}{(x_I^{\tilde g}-1)^2}
-\frac{1}{x_I^{\tilde g}-1}-\frac{(x_J^{\tilde g})^2\log x_J^{\tilde g}}{(x_J^{\tilde g}-1)^2}
+\frac{1}{x_J^{\tilde g}-1}\right ).
\end{align}
\item If $x_I^{\tilde g}=x_J^{\tilde g}$,
\begin{align}
g_{1[1]}(x_I^{\tilde g},x_I^{\tilde g})&=
-\frac{(x_I^{\tilde g}+1)\log x_I^{\tilde g}}{(x_I^{\tilde g}-1)^3}+\frac{2}{(x_I^{\tilde g}-1)^2}~,\nonumber \\
g_{2[1]}(x_I^{\tilde g},x_I^{\tilde g})&=
-\frac{2x_I^{\tilde g}\log x_I^{\tilde g}}{(x_I^{\tilde g}-1)^3}+\frac{x_I^{\tilde g}+1}{(x_I^{\tilde g}-1)^2}~.
\end{align}
\end{itemize}
In this paper, we take $(I,J)=(3,3),~(3,6),~(6,3),~(6,6)$, because we assume the split-family. 
The effective Wilson coefficients are given at the leading order of QCD as follows:
\begin{align}
C_{VLL}(m_b(\Lambda =2~\text{GeV}))=&\eta _{VLL}^{B(K)}C_{VLL}(m_{\tilde g}),\quad 
C_{VRR}(m_b(\Lambda =2~\text{GeV}))=\eta _{VRR}^{B(K)}C_{VLL}(m_{\tilde g}),\nonumber \\
\begin{pmatrix}
C_{SLL}^{(1)}(m_b(\Lambda =2~\text{GeV})) \\
C_{SLL}^{(2)}(m_b(\Lambda =2~\text{GeV}))
\end{pmatrix}&=
\begin{pmatrix}
C_{SLL}^{(1)}(m_{\tilde g}) \\
C_{SLL}^{(2)}(m_{\tilde g})
\end{pmatrix}X_{LL}^{-1}\eta _{LL}^{B(K)}X_{LL},\nonumber \\
\begin{pmatrix}
C_{SRR}^{(1)}(m_b(\Lambda =2~\text{GeV})) \\
C_{SRR}^{(2)}(m_b(\Lambda =2~\text{GeV}))
\end{pmatrix}&=
\begin{pmatrix}
C_{SRR}^{(1)}(m_{\tilde g}) \\
C_{SRR}^{(2)}(m_{\tilde g})
\end{pmatrix}X_{RR}^{-1}\eta _{RR}^{B(K)}X_{RR},\nonumber \\
\begin{pmatrix}
C_{SLR}^{(1)}(m_b(\Lambda =2~\text{GeV})) \\
C_{SLR}^{(2)}(m_b(\Lambda =2~\text{GeV}))
\end{pmatrix}&=
\begin{pmatrix}
C_{SLR}^{(1)}(m_{\tilde g}) \\
C_{SLR}^{(2)}(m_{\tilde g})
\end{pmatrix}X_{LR}^{-1}\eta _{LR}^{B(K)}X_{LR},
\end{align}
where 
\begin{align}
&\eta _{VLL}^B=\eta _{VRR}^B=\left (\frac{\alpha _s(m_{\tilde g})}{\alpha _s(m_t)}\right )^{\frac{6}{21}}
\left (\frac{\alpha _s(m_t)}{\alpha _s(m_b)}\right )^{\frac{6}{23}},\nonumber \\
&\eta _{LL}^B=\eta _{RR}^B=
S_{LL}
\begin{pmatrix}
\eta _{b\tilde g}^{d_{LL}^1} & 0 \\
0 & \eta _{b\tilde g}^{d_{LL}^2}
\end{pmatrix}
S_{LL}^{-1},\qquad 
\eta _{LR}^B=S_{LR}
\begin{pmatrix}
\eta _{b\tilde g}^{d_{LR}^1} & 0 \\
0 & \eta _{b\tilde g}^{d_{LR}^2}
\end{pmatrix}
S_{LR}^{-1},\nonumber \\
&\eta _{b\tilde g}=\left (\frac{\alpha _s(m_{\tilde g})}{\alpha _s(m_t)}\right )^{\frac{1}{14}}
\left (\frac{\alpha _s(m_t)}{\alpha _s(m_b)}\right )^{\frac{3}{46}},\nonumber \\
&\eta _{VLL}^K=\eta _{VRR}^K=\left (\frac{\alpha _s(m_{\tilde g})}{\alpha _s(m_t)}\right )^{\frac{6}{21}}
\left (\frac{\alpha _s(m_t)}{\alpha _s(m_b)}\right )^{\frac{6}{23}}
\left (\frac{\alpha _s(m_b)}{\alpha _s(\Lambda =2~\text{GeV})}\right )^{\frac{6}{25}},\nonumber
\end{align}
\begin{align}
&\eta _{LL}^K=\eta _{RR}^K=
S_{LL}
\begin{pmatrix}
\eta _{\Lambda \tilde g}^{d_{LL}^1} & 0 \\
0 & \eta _{\Lambda \tilde g}^{d_{LL}^2}
\end{pmatrix}
S_{LL}^{-1},\qquad 
\eta _{LR}^K=
S_{LR}
\begin{pmatrix}
\eta _{\Lambda \tilde g}^{d_{LR}^1} & 0 \\
0 & \eta _{\Lambda \tilde g}^{d_{LR}^2}
\end{pmatrix}
S_{LR}^{-1},\nonumber \\
&\eta _{\Lambda \tilde g}=\left (\frac{\alpha _s(m_{\tilde g})}{\alpha _s(m_t)}\right )^{\frac{1}{14}}
\left (\frac{\alpha _s(m_t)}{\alpha _s(m_b)}\right )^{\frac{3}{46}}
\left (\frac{\alpha _s(m_b)}{\alpha _s(\Lambda =2~\text{GeV})}\right )^{\frac{3}{50}},\nonumber \\
&d_{LL}^1=\frac{2}{3}(1-\sqrt{241}),\qquad d_{LL}^2=\frac{2}{3}(1+\sqrt{241}),\qquad 
d_{LR}^1=-16,\qquad d_{LR}^2=2,\nonumber \\
&S_{LL}=
\begin{pmatrix}
\frac{16+\sqrt{241}}{60} & \frac{16-\sqrt{241}}{60} \\
1 & 1
\end{pmatrix},\quad 
S_{LR}=
\begin{pmatrix}
-2 & 1 \\
3 & 0
\end{pmatrix},\nonumber \\
&X_{LL}=X_{RR}=
\begin{pmatrix}
1 & 0 \\
4 & 8
\end{pmatrix},\qquad 
X_{LR}=
\begin{pmatrix}
0 & -2 \\
1 & 0
\end{pmatrix}.\nonumber \\
\end{align}


For the parameters $B_i^{(d)}(i=2-5)$ of $B$ mesons,  we use values in  \cite{Becirevic:2001xt}
as follows:
\begin{eqnarray}
&&B_2^{(B_d)} (m_b)=0.79(2)(4), \qquad
B_3^{(B_d)} (m_b)=0.92(2)(4), \nonumber \\
&&B_4^{(B_d)} (m_b)=1.15(3)(^{+5}_{-7}), \qquad 
B_5^{(B_d)} (m_b)=1.72(4)(^{+20}_{-6}), \nonumber\\
&&B_2^{(B_s)} (m_b)=0.80(1)(4), \qquad
B_3^{(B_s)} (m_b)=0.93(3)(8), \nonumber\\
&&B_4^{(B_s)} (m_b)=1.16(2)(^{+5}_{-7}), \qquad
B_5^{(B_s)} (m_b)=1.75(3)(^{+21}_{-6})\ .
\end{eqnarray}
On the other hand, we use the most updated values for $\hat B_1^{(d)} $ and
 $\hat B_1^{(s)} $ as \cite{Ciuchini-KEKFF2013,Flynn-KEKFF2013}
\begin{equation}
\hat B_1^{(B_s)}  = 1.33\pm 0.06 \ , \qquad  
\hat B_1^{(B_s)} / \hat B_1^{(B_d)}=1.05\pm 0.07 \ . 
\end {equation}

For the paremeters $B_i^{K}(i=2-5)$, we use following values 
  \cite{Allton:1998sm},
\begin{equation}
\begin{split}
B_2^{(K)}(2{\rm GeV})=0.66\pm 0.04 , \qquad 
B_3^{(K)}(2{\rm GeV})=1.05\pm 0.12, \\
B_4^{(K)}(2{\rm GeV})=1.03\pm 0.06, \qquad
B_5^{(K)}(2{\rm GeV})=0.73\pm 0.10,
\end{split}
\end{equation}
and we take recent value of Eq.(\ref{BK}) for deriving $B_1^{(K)}(2{\rm GeV})$.

\section{Squark contribution in $\Delta F=1$ process}

The Wilson coefficients for the gluino contribution 
in Eq.(\ref{hamiltonian}) are written as \cite{GotoNote}

\begin{align}
C_{7\gamma }^{\tilde g}(m_{\tilde g}) &= 
\frac{8}{3}\frac{\sqrt{2}\alpha _s\pi }{2G_FV_{tb}V_{tq}^*} \nonumber \\
&\times \Bigg [\frac{\big (\Gamma _{GL}^{(d)}\big )_{k3}^*}{m_{\tilde d_3}^2}
\left \{ \big (\Gamma _{GL}^{(d)}\big )_{33}\left (-\frac{1}{3}F_2(x_{\tilde g}^3)\right )
+\frac{m_{\tilde g}}{m_b}\big (\Gamma _{GR}^{(d)}\big )_{33}
\left (-\frac{1}{3}F_4(x_{\tilde g}^3)\right )\right \} \nonumber \\
&\hspace{3mm}+\frac{\big (\Gamma _{GL}^{(d)}\big )_{k6}^*}{m_{\tilde d_6}^2}
\left \{ \big (\Gamma _{GL}^{(d)}\big )_{36}\left (-\frac{1}{3}F_2(x_{\tilde g}^6)\right )
+\frac{m_{\tilde g}}{m_b}\big (\Gamma _{GR}^{(d)}\big )_{36}
\left (-\frac{1}{3}F_4(x_{\tilde g}^6)\right )\right \} \Bigg ],
\end{align}
\begin{align}
C_{8G}^{\tilde g}(m_{\tilde g}) &= \frac{8}{3}\frac{\sqrt{2}\alpha _s\pi }{2G_FV_{tb}V_{tq}^*}
\Bigg [\frac{\big (\Gamma _{GL}^{(d)}\big )_{k3}^*}{m_{\tilde d_3}^2}
\left \{ \big (\Gamma _{GL}^{(d)}\big )_{33}
\left (-\frac{9}{8}F_1(x_{\tilde g}^3)-\frac{1}{8}F_2(x_{\tilde g}^3)\right )\right .\nonumber \\
&\hspace{3.8cm}\left .+\frac{m_{\tilde g}}{m_b}\big (\Gamma _{GR}^{(d)}\big )_{33}
\left (-\frac{9}{8}F_3(x_{\tilde g}^3)-\frac{1}{8}F_4(x_{\tilde g}^3)\right )\right \} \nonumber \\
&\hspace{2.4cm}+\frac{\big (\Gamma _{GL}^{(d)}\big )_{k6}^*}{m_{\tilde d_6}^2}
\left \{ \big (\Gamma _{GL}^{(d)}\big )_{36}
\left (-\frac{9}{8}F_1(x_{\tilde g}^6)-\frac{1}{8}F_2(x_{\tilde g}^6)\right )\right .\nonumber \\
&\hspace{3.8cm}\left .+\frac{m_{\tilde g}}{m_b}\big (\Gamma _{GR}^{(d)}\big )_{36}
\left (-\frac{9}{8}F_3(x_{\tilde g}^6)-\frac{1}{8}F_4(x_{\tilde g}^6)\right )\right \} \Bigg ],
\end{align}
where $k=2,1$ correspond to $b\to q~(q=s,d)$ transitions, respectively. 
The loop functions $F_i(x_{\tilde g}^I)$ are given as 
\begin{align}
F_1(x_{\tilde g}^I)&=\frac{x_{\tilde g}^I\log x_{\tilde g}^I}{2(x_{\tilde g}^I-1)^4}
+\frac{(x_{\tilde g}^I)^2-5x_{\tilde g}^I-2}{12(x_{\tilde g}^I-1)^3}~,\nonumber \\
F_2(x_{\tilde g}^I)&=-\frac{(x_{\tilde g}^I)^2\log x_{\tilde g}^I}{2(x_{\tilde g}^I-1)^4}
+\frac{2(x_{\tilde g}^I)^2+5x_{\tilde g}^I-1}{12(x_{\tilde g}^I-1)^3}~,\nonumber \\
F_3(x_{\tilde g}^I)&=\frac{\log x_{\tilde g}^I}{(x_{\tilde g}^I-1)^3}
+\frac{x_{\tilde g}^I-3}{2(x_{\tilde g}^I-1)^2}~,\nonumber \\
F_4(x_{\tilde g}^I)&=-\frac{x_{\tilde g}^I\log x_{\tilde g}^I}{(x_{\tilde g}^I-1)^3}+
\frac{x_{\tilde g}^I+1}{2(x_{\tilde g}^I-1)^2}=\frac{1}{2}g_{2[1]}(x_{\tilde g}^I,x_{\tilde g}^I)~,
\end{align}
with $x_{\tilde g}^I=m_{\tilde g}^2/m_{\tilde d_I}^2~(I=3,6)$.

\section{cEDM}
The cEDM of the strange quark from gluino contribution is given by \cite{GotoNote}
\begin{equation}
d_s^C=-2\sqrt{4\pi \alpha _s(m_{\tilde g})}\text{Im}[A_s^{g22}],
\end{equation}
where 
\begin{align}
A_s^{g22}
&=-\frac{\alpha _s(m_{\tilde g})}{4\pi }\frac{1}{3}
\Bigg [\frac{1}{2m_{\tilde d_3}^2}\bigg \{ 
\Big (m_s(\lambda _{GLL}^{(d)})_3^{22}+m_s(\lambda _{GRR}^{(d)})_3^{22}\Big )
\Big (9F_1(x_{\tilde g}^3)+F_2(x_{\tilde g}^3)\Big )\nonumber \\
&\hspace{3.3cm}+m_{\tilde g}(\lambda _{GLR}^{(d)})_3^{22}
\Big (9F_3(x_{\tilde g}^3)+F_4(x_{\tilde g}^3)\Big )\bigg \} \nonumber \\
&\hspace{-0.7cm}+\frac{1}{2m_{\tilde d_6}^2}\bigg \{ 
\Big (m_s(\lambda _{GLL}^{(d)})_6^{22}+m_s(\lambda _{GRR}^{(d)})_6^{22}\Big )
\Big (9F_1(x_{\tilde g}^6)+F_2(x_{\tilde g}^6)\Big )
+m_{\tilde g}(\lambda _{GLR}^{(d)})_6^{22}
\Big (9F_3(x_{\tilde g}^6)+F_4(x_{\tilde g}^6)\Big )\bigg \} \Bigg ].
\end{align}



\end{document}